\documentclass[aps,prd,twocolumn,nofootinbib,longbibliography,superscriptaddress]{revtex4-1}

\usepackage[utf8]{inputenc}
\usepackage{amsmath,amssymb}
\usepackage{graphicx}
\usepackage{hyperref}
\usepackage{url}
\usepackage{xcolor}
\usepackage{placeins}
\usepackage{balance}
\usepackage{afterpage}
\usepackage{paracol}
\hypersetup{
    colorlinks=true,
    linkcolor=blue,
    citecolor=blue,
    urlcolor=blue
}

\begin{document}

\title{Impact of the SNe Ia Magnitude Transition at 20 Mpc on Cosmological Parameter Estimation}


\author{Leandros Perivolaropoulos}
\email{leandros@uoi.gr}
\affiliation{Department of Physics, University of Ioannina, GR-45110, Ioannina, Greece}

\author{Chrisostomos-Panagiotis Stamou}
\email{c.p.stamou@gmail.com}
\affiliation{Leiden Institute of Physics, Leiden University, The Netherlands}

\date{\today}

\begin{abstract}
We investigate the impact of a late-time transition in the standardized absolute magnitude $M$ on the best-fit values of cosmological parameters using the Pantheon+ dataset. Extending previous analyses which focused on flat $\Lambda$CDM, we examine this transition within flat $\Lambda$CDM, wCDM, and CPL cosmologies, as well as a model-independent cosmographic expansion, employing both frequentist ($\chi^2$ minimization with \textit{AIC}/\textit{BIC}) and Bayesian (MCMC and Nested Sampling) inference frameworks. We confirm that the data consistently favor a step in absolute magnitude of $\Delta M \simeq 0.19~\mathrm{mag}$ at a characteristic distance of $d_{\mathrm{crit}} \approx 20~\mathrm{Mpc}$. The inclusion of this transition leads to a statistically significant improvement in the quality of fit and has a distinct impact on parameter estimation: it induces a systematic increase in the inferred Hubble constant of approximately $2\%$ across all tested models. In contrast, we find that the dynamical parameters governing the background expansion, including the matter density $\Omega_m$ and the dark energy equation of state ($w_0, w_a$), remain stable and largely unaffected. These results indicate that the $20~\mathrm{Mpc}$ feature acts primarily as a low-redshift calibration shift rather than a modification of the late-time expansion history.
\end{abstract}

\maketitle

\section{Introduction}

A major challenge in contemporary cosmology is the persistent discrepancy between the value of the Hubble constant \(H_0\) inferred from early- and late-universe observations, commonly referred to as the Hubble tension or Hubble crisis\cite{Perivolaropoulos:2021jda,Verde:2019ivm,Bernal:2016gxb,Huterer:2017buf}. In the context of a flat \(\Lambda\)CDM model, analysis of Cosmic Microwave Background (CMB) anisotropies from the \textit{Planck} satellite yields \(H_0 = 67.4 \pm 0.5 \ \mathrm{km \ s^{-1} \ Mpc^{-1}}\) \cite{Planck:2018vyg}, while recent local measurements using the Pantheon+ Type Ia supernova (SNe Ia) compilation calibrated with Cepheid variable stars give \(H_0 = 73.6 \pm 1.4 \ \mathrm{km \ s^{-1} \ Mpc^{-1}}\) \cite{Riess:2021jrx}. The tension between these determinations exceeds \(5\sigma\), suggesting the possible presence of unaccounted systematics or new physics beyond \(\Lambda\)CDM.

The Pantheon+ compilation \cite{Brout:2022vxf} contains 1701 light curves corresponding to 1550 distinct SNe Ia in the redshift range \(0.001 < z < 2.26\). Of these, 42 SNe are hosted in galaxies with Cepheid distance measurements, providing an absolute calibration of the standardized magnitude \(M\) and breaking the degeneracy between \(H_0\) and \(M\), which allows a direct estimation of $H_{0}$ via the distance ladder method. In this approach, Cepheid distances anchor nearby SNe Ia, which in turn calibrate more distant events, extending the Hubble diagram to cosmological scales \cite{Scolnic:2017caz,Dhawan:2020xmp,Kessler_2017,Riess:2016jrr}. In contrast, the CMB-inferred value of \(H_0\) is obtained by fitting the angular power spectrum of temperature and polarization anisotropies with a flat \(\Lambda\)CDM model, where \(H_0\) is a derived parameter constrained primarily by the angular scale of the sound horizon at recombination \cite{Aghanim:2018eyx,Akrami:2018vks,Addison:2015wyg,Knox:2019rjx,Riess:2019cxk}.

The reliability of the Pantheon+ sample has been extensively examined in the literature, with studies focusing on its calibration methods, statistical homogeneity, and sensitivity to astrophysical systematics \cite{Scolnic:2021amr,Carr_2022,Peterson_2022,hu2023testingcosmologicalprinciplepantheon,Dhawan:2023ekc,Colgain:2022tql,Wang:2022ssr,Kenworthy:2022jdh}. Various works have explored possible explanations for the Hubble tension \cite{Perivolaropoulos:2021jda,DiValentino:2021izs}, including local large-scale structure effects \cite{Camarena_2020,huterer2023enoughlocalvoidsolve,Sola:2017znb}, calibration uncertainties \cite{Efstathiou_2021,Khetan:2020hmh,Scolnic_2014,lu2023supernovacalibrationgravitationalwave}, and extensions of \(\Lambda\)CDM such as early dark energy, interacting dark sectors, and modifications to the SNe Ia standardization process\cite{Poulin:2018cxd,poulin2023upsdownsearlydark,Kamionkowski:2022pkx,Buen-Abad:2017gxg,buenabad2024atomicdarkmatterinteracting,efstathiou2023improvedplanckconstraintsaxionlike,Rigault_2020,popovic2024modellingimpacthostgalaxy,toy2025reductiontypeiasupernova,Poulin_2018,Aylor_2019,Cai_2022,bansal2026difficultieslatetimesolutionshubble}.

Motivated by possible inhomogeneities in SNe Ia luminosities\cite{Perivolaropoulos:2021bds,Marra:2021fvf,Alestas:2021nmi}, the authors of \cite{Perivolaropoulos_2023} proposed a model in which the absolute magnitude \(M\) undergoes a discrete transition at a critical luminosity distance \(d_{\mathrm{crit}}\), separating supernovae into a low-distance sample with \(M_<\) and a high-distance sample with \(M_>\). In their baseline analysis, using a standard $\chi^{2}$ minimization, without assuming any transition, they inferred \(M = -19.253 \pm 0.03\), \(H_0 = 73.4 \pm 1.1 \ \mathrm{km \ s^{-1} \ Mpc^{-1}}\), and \(\Omega_m = 0.333 \pm 0.018\), in excellent agreement with the Pantheon+ results. When allowing for a transition and letting the data determine \(d_{\mathrm{crit}}\), they found a best-fit value of \(d_{\mathrm{crit}} \approx 20 \ \mathrm{Mpc}\), with \(M_<\) and \(M_>\) differing by \(\Delta M \approx 0.187\) mag. This change corresponded to an increase of approximately \(2\%\) in the inferred \(H_0\) relative to the no-transition case, yielding \(H_0 = 74.9 \pm 1.1 \ \mathrm{km \ s^{-1} \ Mpc^{-1}}\). The statistical preference for the transition model in a flat \(\Lambda\)CDM background reached \(\sim 3\sigma\), and an Akaike Information Criterion (AIC) \cite{1100705} analysis indicated that the transition model provided a better fit to the Pantheon+ data than the standard no-transition case.

A potential systematic affecting low-redshift SNe Ia analyses is the \textit{volumetric redshift scatter bias} \cite{Kenworthy:2022jdh,Brout:2022vxf}, which arises when peculiar velocities and redshift measurement errors, combined with a flux-limited survey, lead to an overrepresentation of intrinsically brighter objects at very low redshifts. \cite{Perivolaropoulos_2023} examined this effect by removing all non-Cepheid-hosted SNe Ia with \(z < 0.01\) from their sample. They found that the significance of the \(M\) transition decreased from \(\sim 3\sigma\) to \(\sim 2\sigma\), but the preference for a transition near 20 Mpc persisted, also validated by an AIC test, which now mildly preferred the transition model instead of strongly as in the original case, including all SNe.

An open question is how such a luminosity transition would affect cosmological inferences in scenarios beyond flat \(\Lambda\)CDM \cite{Benevento:2020fev,DiValentino:2017rcr,Banihashemi:2018oxo,Alestas:2021luu}. In particular: 
\begin{itemize}
\item How would a dynamical dark energy background, modeled via wCDM or CPL parametrizations, respond to such a feature? 
\item How would the transition manifest in a cosmographic expansion of the luminosity distance, independent of a specific cosmological model?
\end{itemize}

In this work, we address these questions by first verifying the flat \(\Lambda\)CDM results of \cite{Perivolaropoulos_2023} using both frequentist $\chi^2$ minimization and Bayesian inference with Markov Chain Monte Carlo (MCMC) \cite{1953JChPh..21.1087M,1970Bimka..57...97H} and Nested Sampling \cite{2004AIPC..735..395S}. We then extend the analysis to a second-order cosmographic expansion, wCDM, and CPL models \cite{CHEVALLIER_2001,Linder_2003}, assessing the significance and robustness of the inferred \(M\) transition across these frameworks. 

Two Bayesian sampling methods are used in a complementary way. MCMC efficiently explores the posterior distribution given a specified likelihood and priors, but is blind to the Bayesian evidence \(\mathcal{Z}\). This makes it less computationally expensive and well suited for obtaining statistically robust posterior samples and parameter estimates. In our analysis, we employ 32 walkers with 5000 steps each, discarding the first \(20\%\) as burn-in, which yields stable posteriors across all tested models. 

Nested Sampling, by contrast, simultaneously produces posterior samples and an accurate estimate of the Bayesian evidence, enabling quantitative model comparison. The trade-off is higher computational cost. To mitigate this, we use a reduced number of live points, which increases statistical uncertainty in the posterior but still provides reliable evidence estimates. As shown in Section~\ref{Sec.III}, parameter constraints from Nested Sampling remain consistent with those from the higher-statistics MCMC runs, validating this combined approach. Finally, we note that MCMC is less sensitive than direct $\chi^2$ minimization to models with many free parameters. In particular, while $\chi^2$ minimization works well for most cases considered here, it fails to yield stable results for the transition CPL model, which involves seven parameters. For this case, the analysis is performed exclusively within the Bayesian framework.

The paper is organized as follows. Section~\ref{Sec.II} describes the relevant Pantheon+ data used in this work, defines the theoretical framework for parameter inference,the cosmological models and the cosmographic expansion considered and specifies their corresponding Hubble expansion rates \(H(z)\), and explains the implementation of the \(M\) transition in each case. We first reproduce and extend the results of \cite{Perivolaropoulos_2023} using their standard $\chi^2$ minimization and a Bayesian analysis via MCMC sampling and Nested Sampling, comparing the original and transition \(\Lambda\)CDM models. This serves as the baseline for assessing the impact of the transition on the cosmographic expansion and on dynamical dark energy models. Section~\ref{Sec.III} presents the results, including confidence contours for the frequentist analysis and credible intervals for the Bayesian analysis. Finally, Section~\ref{Sec.IV} summarizes our findings and discusses their implications.

\section{Data, Methods and Cosmological Framework}\label{Sec.II}

\subsection{Pantheon+ Dataset}\label{IIA}

Our analysis is based on the Pantheon+ compilation \cite{Brout:2022vxf}, which contains 1701 light curves corresponding to 1550 distinct Type Ia supernovae (SNe Ia) in the redshift range \(0.001 < z < 2.26\). Of these, 42 unique SNe are hosted in galaxies with Cepheid distance measurements, providing an absolute calibration of the standardized magnitude \(M\) and breaking the degeneracy between \(H_0\) and \(M\) \cite{Guy_2007}.

From the publicly available Pantheon+ data file, we use the following columns:
\begin{itemize}
    \item Column 3: Hubble diagram redshift in the CMB frame (\(z_{\mathrm{HD}}\)), corrected for the Solar System’s peculiar motion.
    \item Column 9: Corrected apparent $B$-band magnitude (\(m_B\)).
    \item Column 10: Statistical uncertainty in \(m_B\).
    \item Column 13: Cepheid-calibrated distance modulus (\(\mu_{\mathrm{Cepheid}}\)), with a value of $-9$ for non-Cepheid-hosted SNe.
    \item Column 14: Binary flag identifying whether a SN is Cepheid-hosted (1) or not (0).
\end{itemize}
The analysis incorporates the full $1701 \times 1701$ covariance matrix provided by the Pantheon+ team, which includes both statistical and systematic uncertainties as well as correlations between all light curves.

\subsection{Statistical Methodology}\label{SecII.Methods}

\subsubsection{Frequentist Inference: $\chi^2$ Minimization}\label{SecII.B}

We first perform a frequentist analysis by minimizing the $\chi^2$ likelihood
\begin{equation}
\chi^2(\boldsymbol{\theta}) = \mathbf{Q}^T \, C^{-1} \, \mathbf{Q} \, ,
\label{eq1}
\end{equation}
where $\mathbf{Q} = \boldsymbol{\mu}_{\mathrm{obs}} - \boldsymbol{\mu}_{\mathrm{th}}(\boldsymbol{\theta})$ is the vector of residuals between the observed and theoretical distance moduli, $\boldsymbol{\theta}$ denotes the set of cosmological parameters, and $C$ is the total covariance matrix including both statistical and systematic uncertainties as provided in the Pantheon+ dataset. The theoretical distance modulus is defined as
\begin{equation}
\mu_{\mathrm{th}}(z;\boldsymbol{\theta}) = 5 \log_{10} \left( \frac{d_L(z;\boldsymbol{\theta})}{\mathrm{Mpc}} \right) + 25 \, ,
\label{eq2}
\end{equation}
where $d_L(z;\boldsymbol{\theta})$ is the luminosity distance. In a flat Friedmann–Lemaître–Robertson–Walker (FLRW) cosmology, the luminosity distance is related to the Hubble expansion rate $H(z)$ by
\begin{equation}
d_L(z;\boldsymbol{\theta}) = (1+z) \, c \int_0^z \frac{dz'}{H(z';\boldsymbol{\theta})} \, .
\label{eq3}
\end{equation}
For example, in the flat $\Lambda$CDM model,
\begin{equation}
H(z) = H_0 \, \sqrt{\Omega_m (1+z)^3 + 1 - \Omega_m} \, ,
\label{eq4}
\end{equation}
so that $d_L(z)$ follows directly from \eqref{eq3} and \eqref{eq4}. The minimization of \eqref{eq1} with respect to $\boldsymbol{\theta}$ yields the best-fit parameter values and the value of the $\chi^2$ at minimum, $\chi^{2}_{min}$. The goodness of fit is quantified by the reduced $\chi^2$, defined as:
\begin{equation}\label{eq5}
\chi^2_{\mathrm{red}} = \frac{\chi^2_{\mathrm{min}}}{N-k},
\end{equation}
where $N$ is the number of data points and $k$ is the number of free parameters. The Akaike Information Criterion (AIC) and the Bayesian Information Criterion (BIC) \cite{1978AnSta...6..461S,Liddle_2004,Trotta_2008} are used to assess the relative quality of different models while penalizing model complexity. Definition and interpretation of AIC/BIC values are reported in Appendix and in Table \ref{tab:aicbic}.

The statistical uncertainty on the best-fit parameters is estimated from the curvature of the $\chi^2$ surface at its minimum. This is quantified through the \textit{Fisher information matrix}, defined as
\begin{equation}\label{eq8}
F_{ij} = \frac{1}{2} \, \frac{\partial^2 \chi^2}{\partial \theta_i \, \partial \theta_j} \bigg|_{\boldsymbol{\theta} = \boldsymbol{\theta}_{\mathrm{bf}}} \, ,
\end{equation}
where $\boldsymbol{\theta}_{\mathrm{bf}}$ denotes the best-fit parameter set. The inverse of the Fisher matrix, $F^{-1}$, provides the parameter covariance matrix, whose diagonal elements correspond to the variances of each parameter. The square root of these diagonal elements gives the $1\sigma$ uncertainties reported for the inferred parameters.

\subsubsection{Bayesian Inference}\label{SecII.C}

In the Bayesian framework, the inference of model parameters $\boldsymbol{\theta}$ given data $\mathbf{d}$ is based on Bayes’ theorem,
\begin{equation}\label{eq9}
P(\boldsymbol{\theta} | \mathbf{d}) = \frac{\mathcal{L}(\mathbf{d} | \boldsymbol{\theta}) \, \pi(\boldsymbol{\theta})}{\mathcal{Z}} \, ,
\end{equation}
where $\pi(\boldsymbol{\theta})$ is the prior probability density of the parameters, $\mathcal{L}(\mathbf{d} | \boldsymbol{\theta})$ is the likelihood function, and
\begin{equation}\label{eq10}
\mathcal{Z} = \int \mathcal{L}(\mathbf{d} | \boldsymbol{\theta}) \, \pi(\boldsymbol{\theta}) \, d\boldsymbol{\theta}
\end{equation}
is the Bayesian evidence, which acts as a normalization constant in parameter estimation and enables quantitative model comparison.

In our case, $\mathbf{d}$ corresponds to the Pantheon+ measurements, consisting of the redshifts, corrected apparent magnitudes of Type Ia supernovae and Cepheid distance moduli. The likelihood is assumed Gaussian in the residual vector $\mathbf{Q} = \mu_{\mathrm{obs}} - \mu_{\mathrm{th}}(\boldsymbol{\theta})$:
\begin{equation}\label{eq11}
\mathcal{L}(\mathbf{d} | \boldsymbol{\theta}) \propto \exp\left( -\frac{1}{2} \mathbf{Q}^T C^{-1} \mathbf{Q} \right) ,
\end{equation}
with $\mu_{\mathrm{th}}(\boldsymbol{\theta})$ computed from the luminosity distance corresponding to the chosen cosmological model. For instance, in the flat $\Lambda$CDM case, $\boldsymbol{\theta} = \{M, H_0, \Omega_m\}$. Once the posterior distribution $P(\boldsymbol{\theta} | \mathbf{d})$ is sampled, the probability distributions of each of the parameters can be found via marginalization: 
\begin{equation}\label{eq12}
P(\theta_i | \mathbf{d}) = \int P(\boldsymbol{\theta} | \mathbf{d}) \, d\theta_{j\neq i} \, ,
\end{equation}
From these marginalized distributions, medians, means, and MAP values of the parameters can be computed to, assess the symmetry, unimodality, and possible skewness of the posterior distributions. In our work, we adopt flat (uniform) priors within physically motivated ranges, and we report posterior mean values following Pantheon+. The adopted prior ranges are summarized in Appendix. Beyond parameter estimation, the Bayesian evidence $\mathcal{Z}$ serves as a quantitative measure for model comparison. Given two competing models, $M_1$ and $M_2$, with evidences $\mathcal{Z}_1$ and $\mathcal{Z}_2$, their relative preference is expressed through the \textit{Bayes factor},
\begin{equation}\label{eq13}
B_{12} = \frac{\mathcal{Z}_1}{\mathcal{Z}_2} \, ,
\end{equation}
or equivalently in logarithmic form as $\Delta \log \mathcal{Z} = \log \mathcal{Z}_1 - \log \mathcal{Z}_2$. A positive value of $\Delta \log \mathcal{Z}$ indicates preference for Model~1, while a negative value favors Model~2. The strength of this preference is commonly interpreted according to the \textit{Jeffreys scale} \cite{Robert_2009}, which provides a guideline for assessing the degree of evidence in favor of one model over another. The interpretation used in this work is summarized in Table~\ref{tab:jeffreys} in the Appendix.

\subsubsection{MCMC \& Nested Sampling Setup}\label{SecII.D}

While MCMC is well suited for parameter estimation, it does not directly provide the Bayesian evidence $\mathcal{Z}$ required for model comparison. For this purpose, we employ the \texttt{dynesty} nested sampling algorithm \cite{Speagle:2019ivv}, which simultaneously yields posterior samples and an accurate estimate of $\mathcal{Z}$. The ratio of Bayesian evidences between two models defines the Bayes factor, which quantifies the relative preference for one model over another.

\paragraph{\underline{\textbf{MCMC:}}}
We use the affine-invariant ensemble sampler implemented in \texttt{emcee} \cite{Foreman_Mackey_2013}. The sampler across all models is initialized with 32 walkers, each running for 5,000 steps, with the first 1,000 steps discarded as burn-in to ensure that the chains have reached the equilibrium distribution. Convergence is evaluated through the integrated autocorrelation time, which measures the number of steps required for successive samples to become effectively uncorrelated. From these values we compute the effective sample size (ESS), quantifying how many independent samples contribute to the posterior statistics. Well-mixed chains are characterized by short autocorrelation times and large ESS values, ensuring that the inference is based on a large number of effectively uncorrelated samples. These diagnostics confirm that the MCMC posteriors are statistically reliable across all tested models.

\paragraph{\underline{\textbf{Nested Sampling:}}}
For the computation of the Bayesian evidence and direct model comparison, we use the \texttt{dynesty} nested sampling algorithm \cite{Speagle_2020}. The sampler is configured with 500 live points across all models, besides CPL for which 250 live points were used given the model's complexity, the \texttt{rwalk} sampling method (a random-walk-based proposal), and the \texttt{multi} bounding option to efficiently enclose the evolving live-point cloud. Parallel execution is enabled through a multiprocessing pool to accelerate convergence. Given computational constraints, we could not exceed 500 live points; as a result, the corresponding contour plots appear slightly less smooth than those obtained from MCMC chains, especially for models with additional free parameters such as the transition model. Nevertheless, this configuration is sufficient for robust computation of the Bayesian evidence and relative model comparison, provided that the same number of live points is used consistently across all tested models.

Convergence of the nested sampling runs is assessed using the internal \texttt{dynesty} stopping criterion based on the remaining evidence contribution, $\Delta \log \mathcal{Z}$. Typical termination thresholds of $\Delta \log \mathcal{Z} < 0.5$ guarantee that the evidence integral has stabilized. Additional indicators, such as the total number of iterations, likelihood evaluations, and sampling efficiency (typically a few percent for moderate-dimensional problems), provide complementary confirmation that the posterior space has been adequately explored. Together, these diagnostics ensure that both MCMC and Nested Sampling analyses yield consistent, converged, and statistically robust results. As a representative example of our convergence diagnostics, for the most parameter-rich case (transition CPL) the integrated autocorrelation times are
$\tau \simeq (83\text{--}227)$ steps across parameters, corresponding to effective sample sizes $N_{\rm eff}\simeq (7\times10^2\text{--}2\times10^3)$ for our post-burn-in chains (32 walkers $\times$ 4000 steps).
For Nested Sampling (transition CPL), a typical run terminates with $\Delta\log\mathcal{Z}\ll 0.5$ (e.g.\ $\Delta\log\mathcal{Z}\simeq 0.001$) and sampling efficiency of a few percent (e.g.\ $\sim3.8\%$), yielding $\log\mathcal{Z}=-769.66\pm0.25$ and an effective posterior sample size ${\rm ESS}\simeq 1666$.
Analogous diagnostics were verified across the remaining models and were found to be consistent with well-mixed chains and stabilized evidence estimates.

\subsubsection{Cepheid Calibrator and Likelihood}\label{SecII.E}

The observed apparent magnitude $m_{B,i}$ of a Type~Ia supernova is related to its absolute magnitude $M$ and luminosity distance $d_{L,i}$ through the standard distance–modulus relation,
\begin{equation}\label{eq14}
\mu_i = m_{B,i} - M = 5 \log_{10} \!\left( \frac{d_{L,i}}{\mathrm{Mpc}} \right) + 25 .
\end{equation}
In cosmological analyses based solely on Hubble–flow supernovae, the parameters $M$ and $H_0$ are fully degenerate, since an overall shift in $M$ can be compensated by a rescaling of $H_0$ in the luminosity distance, as can be seen via equations \eqref{eq3},\eqref{eq4}. Consequently, the two quantities cannot be determined independently without an external calibration. This degeneracy is broken by including the subset of Cepheid–hosted supernovae, for which the absolute distance modulus $\mu_{\mathrm{Ceph},i}$ has been independently measured using the SH0ES distance ladder. For these objects, the absolute magnitude can be directly inferred as $M = m_{B,i} - \mu_{\mathrm{Ceph},i}$, thereby anchoring the distance scale and allowing the simultaneous inference of both $M$ and $H_0$ within the same likelihood framework.

The incorporation of Cepheid–hosted SNe modifies the construction of the residual vector $\mathbf{Q}$ that enters the $\chi^2$ statistic. For the standard (no–transition) model, the components of $\mathbf{Q}$ are defined as
\begin{equation}
Q_i = 
\begin{cases}
m_{B,i} - M - \mu_{\mathrm{Ceph},i} & \text{if SN $i$ is in a Cepheid host,} \\
m_{B,i} - M - \mu_{\mathrm{th}}(z_i) & \text{otherwise}.
\end{cases}
\label{eq15}
\end{equation}
where $\mu_{\mathrm{th}}(z_i)$ expresses the model under consideration with parameters $\vec{\theta}$ and is given by:
\begin{equation}\label{eq16}
\mu_{\mathrm{th}}(z;\vec{\theta}) = 5\log_{10}\!\left(\frac{d_L(z;\vec{\theta})}{\mathrm{Mpc}}\right) + 25 \, .
\end{equation}
This formulation ensures that Cepheid–hosted SNe directly constrain the absolute magnitude, while the remaining objects constrain the cosmological parameters through their redshift–dependent model distances.

When a possible luminosity transition is introduced at a critical distance $d_{\mathrm{crit}}$, corresponding to a distance modulus $\mu_{\mathrm{crit}}$, the absolute magnitude is allowed to take two distinct values, $M_{<}$ and $M_{>}$, below and above the transition scale, respectively. 

\vspace{0.3cm}
\noindent
The residual vector then generalizes to
\begin{equation}\label{eq17}
\hspace{-0.2cm}
Q_i =
\begin{cases}
m_{B,i} - M_{<} - \mu_{\mathrm{Ceph},i} & \text{if } \mu_{i,S} < \mu_{\mathrm{crit}},~i \in \text{Cepheid hosts}, \\[4pt]
m_{B,i} - M_{>} - \mu_{\mathrm{Ceph},i} & \text{if } \mu_{i,S} > \mu_{\mathrm{crit}},~i \in \text{Cepheid hosts}, \\[4pt]
m_{B,i} - M_{<} - \mu_{\mathrm{th}}(z_i) & \text{if } \mu_{i,S} < \mu_{\mathrm{crit}},~i \notin \text{Cepheid hosts}, \\[4pt]
m_{B,i} - M_{>} - \mu_{\mathrm{th}}(z_i) & \text{if }\mu_{i,S} > \mu_{\mathrm{crit}},~i \notin \text{Cepheid hosts},
\end{cases}
\end{equation}
\vspace{0.2cm}

\noindent
where $\mu_{i,S} = m_{B,i} + 19.253$ with $M_{\rm SH0ES}\equiv -19.253$ being the SH0ES best-fit absolute magnitude used as a reference offset \cite{Riess:2021jrx}, and $\mu_{\mathrm{crit}}$ corresponds to the distance modulus at the critical distance $d_{\mathrm{crit}}$. This extended likelihood model allows testing for deviations from homogeneity in the standardized SNe Ia luminosity and will be used for each of the models that will be tested.

\subsection{Cosmological Models}\label{SecII.Models}

We test four background models, each defined by a specific expansion history \(H(z)\) and a set of free parameters \(\vec{\theta}\). The luminosity distance is computed from Eq.~(\ref{eq3}) in all cases besides the cosmographic expansion, for which the luminosity distance is calculated via a Taylor expansion and it's therefore model-independent. The full parameter set includes the absolute-magnitude parameter(s) $M$ (no-transition) or ${M_<,M_>}$ (transition), jointly constrained with the background parameters \(\vec{\theta}\). Finally, the convergence criteria for the MCMC and Nested Sampling runs are satisfied.

\begin{itemize}
    \item \textbf{Flat $\Lambda$CDM} \\
    Parameter vector: \(\vec{\theta} = \{H_0,\,\Omega_m\}\) \\
    The Hubble parameter is given by Eq.~(\ref{eq4}), while \(M\) is treated as a nuisance parameter determined through calibration with the Cepheid–anchored SNe.

    \item \textbf{Cosmographic Expansion (2nd order)} \\
    Parameter vector: \(\vec{\theta} = \{H_0,\,q_0\}\) \\
    The cosmographic model provides a purely kinematic description of the late–time expansion, independent of any cosmological assumptions. \cite{Visser_2004,Xia_2012,Catto_n_2007} Expanding the luminosity distance up to second order in redshift yields
    \begin{equation}\label{eq18}
    d_L(z) = \frac{c}{H_0}\left[z + \frac{1}{2}(1 - q_0) z^2\right] .
    \end{equation}
    This expression is valid for low redshifts of the order \(z \lesssim 0.15\) and is used to probe the local expansion rate and its possible deviations from homogeneity. The choice of the cut-off redshift scale is explained in Appendix.

    \item \textbf{Flat $w$CDM} \\
    Parameter vector: \(\vec{\theta} = \{H_0,\,\Omega_m,\,w\}\) \\
    The expansion history is governed by
    \begin{equation}\label{eq19}
    H(z) = H_0\,\sqrt{ \Omega_m (1+z)^3 + (1-\Omega_m)(1+z)^{3(1+w)} } \, .
    \end{equation}

    \item \textbf{Flat CPL ($w_0w_a$CDM)} \\
    Parameter vector: \(\vec{\theta} = \{H_0,\,\Omega_m,\,w_0,\,w_a\}\) \\
    For a time–varying dark–energy equation of state \(w(z) = w_0 + w_a\,z/(1+z)\), the Hubble parameter becomes
    \begin{multline}\label{eq20}
    H(z) = H_0 \Big[\Omega_m(1+z)^3 \\
    + (1-\Omega_m)(1+z)^{3(1+w_0+w_a)} e^{-3w_a z/(1+z)}\Big]^{1/2}\! .
    \end{multline}
\end{itemize}

\section{Results}\label{Sec.III}

\subsection{Flat \texorpdfstring{$\Lambda$}{Lambda}CDM}\label{sec:III_LCDM}
We begin with the flat $\Lambda$CDM baseline analyzed by \cite{Perivolaropoulos_2023}, who used $\chi^2$ minimization in a frequentist framework to test for a late-time transition in the standardized absolute magnitude $M$. We independently reproduced their frequentist analysis and additionally performed a Bayesian inference (MCMC and Nested Sampling). Across methods, our results are in excellent agreement with \cite{Perivolaropoulos_2023}: the transition model is favored by information criteria (negative $\Delta$AIC and $\Delta$BIC) and by the Bayes factor (positive $\Delta\log\mathcal{Z}$). The critical distance is inferred directly from the data and is consistently found to be $d_{\rm crit}\simeq 20~\mathrm{Mpc}$, indicating a low-redshift, late-time transition. Moreover, $\Omega_m$ remains essentially unchanged between models, while $H_0$ increases by approximately $2\%$ in the transition case. The complete parameter constraints are reported in Table~\ref{tab:lcdm_all}, and the corresponding posterior/corner and frequentist contour plots (1, 2, 3$\sigma$) are shown in Figs.~\ref{fig:lcdm_cosmo_contours},\ref{fig:corner_lcdm_noT},\ref{fig:corner_lcdm_T}.

\subsection{Cosmographic Expansion (2nd order)}\label{sec:III_Cosmography}
For the cosmographic model truncated at $\mathcal{O}(z^2)$, a redshift cut-off is required to ensure validity of the series while retaining sufficient SNe to constrain the deceleration parameter $q_0$ with reasonable uncertainty. We adopt $z_{\max}=0.15$ and verify this choice by scanning $z_{\max}\in\{0.05,\,0.10,\,0.15,\,0.18\}$, finding that $z_{\max}=0.15$ provides the optimal balance: very low cuts ($0.05$) are dominated by local scatter and yield large errors, intermediate cuts ($0.10$) leave $q_0$ weakly constrained, while extending to $0.18$ begins to require higher-order terms, degrading the constraint. Details are provided in Appendix. Using $z_{\max}=0.15$, we again find that the transition model is supported by the data (improved $\chi^2$ with negative $\Delta$AIC/$\Delta$BIC and positive $\Delta\log\mathcal{Z}$), with a late-time transition at $d_{\rm crit}\simeq 20~\mathrm{Mpc}$. The full results are listed in Table~\ref{tab:cosmo_all}, and the posterior/corner and frequentist contour plots (1, 2, 3$\sigma$) are shown in Appendix~\ref{app:suppfigs}.

\subsection{Flat \texorpdfstring{$w$}{w}CDM}\label{sec:III_wCDM}
Allowing dynamical dark energy with a constant equation of state $w$ (flat $w$CDM), we repeat the parallel frequentist and Bayesian analyses used for $\Lambda$CDM. The qualitative picture persists across methods: the data prefer a transition at $d_{\rm crit}\sim 20~\mathrm{Mpc}$, while $\Omega_m$ and $w$ remain largely unaffected by including the transition; $H_0$ increases in the transition case by an amount comparable to the $\Lambda$CDM result. These outcomes are fully consistent with the interpretation that the transition chiefly
\balance
impacts the absolute calibration while leaving the background evolution parameters broadly stable. Complete constraints appear in Table~\ref{tab:wcdm_all}, with the frequentist contour plots shown in Figs.~\ref{fig:wcdm_overlay},\ref{fig:wcdm_contours} and the corresponding Bayesian corner plots shown in Appendix~\ref{app:suppfigs}.

\subsection{Flat CPL (\texorpdfstring{$w_0w_a$}{w0wa}CDM)}\label{sec:III_CPL}
For the CPL parameterization, $w(z)=w_0+w_a\,z/(1+z)$, the higher dimensionality and known instability of direct $\chi^2$ minimization render a purely Bayesian approach preferable; we therefore perform MCMC and Nested Sampling only. The results follow the same pattern as above: the transition model is supported with $d_{\rm crit}\sim 20~\mathrm{Mpc}$, $H_0$ increases relative to the no-transition case, and $(w_0,w_a)$ remain consistent within uncertainties with the no-transition constraints. The full constraints are summarized in Table~\ref{tab:cpl_all}, and the Bayesian corner plots are shown in Appendix~\ref{app:suppfigs}
.
\nobalance
\FloatBarrier
\begin{widetext}

\begin{table*}
\centering
\caption{\textbf{Flat $\Lambda$CDM — Frequentist best-fit $\pm1\sigma$ and Bayesian posterior means $\pm$ 68\% CIs} for no–transition and transition. For Bayesian evidence, $\Delta\log\mathcal{Z}$ is transition $-$ no–transition.}
\label{tab:lcdm_all}
\begin{tabular}{l|ccc|ccc}
\hline\hline
 & \multicolumn{3}{c|}{\textbf{No transition}} & \multicolumn{3}{c}{\textbf{Transition}} \\
\cline{2-7}
\textbf{Parameter} & \textbf{Frequentist} & \textbf{Bayes (MCMC)} & \textbf{Bayes (Nested)} & \textbf{Frequentist} & \textbf{Bayes (MCMC)} & \textbf{Bayes (Nested)} \\
\hline
$M$ & $-19.25 \pm 0.03$ & $-19.249 \pm 0.029$ & $-19.247 \pm 0.030$ & --- & --- & --- \\
$M_<$ & --- & --- & --- & $-19.398 \pm 0.05$ & $-19.404 \pm 0.056$ & $-19.403 \pm 0.056$ \\
$M_>$ & --- & --- & --- & $-19.206 \pm 0.03$ & $-19.213 \pm 0.031$ & $-19.213 \pm 0.031$ \\
$\Delta M$ & --- & --- & --- & $0.192$ & $0.191$ & $0.190$ \\
$H_0$ [km s$^{-1}$ Mpc$^{-1}$] & $73.42 \pm 1.01$ & $73.41 \pm 1.00$ & $73.46 \pm 1.02$ & $74.80 \pm 1.01$ & $74.62 \pm 1.08$ & $74.61 \pm 1.08$ \\
$\Omega_m$ & $0.333 \pm 0.018$ & $0.333 \pm 0.018$ & $0.333 \pm 0.018$ & $0.332 \pm 0.018$ & $0.333 \pm 0.018$ & $0.333 \pm 0.018$ \\
$d_{\rm crit}$ [Mpc] & --- & --- & --- & $19.95 \pm 0.1$ & $19.66 \pm 0.95$ & $19.69 \pm 1.01$ \\
\hline
$\chi^2_{\min}$ & 1522.98 & --- & --- & 1503.25 & --- & --- \\
$\chi^2_{\rm red}$ & 0.90 & --- & --- & 0.89 & --- & --- \\
$\Delta\chi^2$ (vs no-trans) & --- & --- & --- & $-19.73$ & --- & --- \\
$\Delta$AIC & --- & --- & --- & $-15.73$ & --- & --- \\
$\Delta$BIC & --- & --- & --- & $-4.85$ & --- & --- \\
\hline
\textbf{log$\boldsymbol{\mathcal{Z}}$ (Nested)} & --- & --- & $-771.45 \pm 0.24$ & --- & --- & $-767.49 \pm 0.28$ \\
\hline
\multicolumn{7}{c}{$\Delta\log\mathcal{Z}$ (transition $-$ no–transition): \ \ +3.96 $\pm$ 0.37} \\
\hline
\multicolumn{7}{c}{\textit{Bayesian Interpretation:} Moderate evidence for transition model} \\
\hline
\multicolumn{7}{c}{\textit{AIC Interpretation:} Strong preference for transition model \;|\; 
\textit{BIC Interpretation:} Weak preference for transition model} \\
\hline\hline
\end{tabular}
\end{table*}

\begin{table*}[t]
\centering
\caption{\textbf{Cosmographic Expansion (2nd order) — Frequentist best-fit $\pm1\sigma$ and Bayesian posterior means $\pm$ 68\% CIs} for no–transition and transition (with $z_{\max}=0.15$). For Bayesian evidence, $\Delta\log\mathcal{Z}$ is transition $-$ no–transition.}
\label{tab:cosmo_all}
\begin{tabular}{l|ccc|ccc}
\hline\hline
 & \multicolumn{3}{c|}{\textbf{No transition}} & \multicolumn{3}{c}{\textbf{Transition}} \\
\cline{2-7}
\textbf{Parameter} & \textbf{Frequentist} & \textbf{Bayes (MCMC)} & \textbf{Bayes (Nested)} & \textbf{Frequentist} & \textbf{Bayes (MCMC)} & \textbf{Bayes (Nested)} \\
\hline
$M$ & $-19.25 \pm 0.03$ & $-19.248 \pm 0.029$ & $-19.248 \pm 0.029$ & --- & --- & --- \\
$M_<$ & --- & --- & --- & $-19.401 \pm 0.05$ & $-19.405 \pm 0.055$ & $-19.409 \pm 0.055$ \\
$M_>$ & --- & --- & --- & $-19.207 \pm 0.03$ & $-19.212 \pm 0.031$ & $-19.213 \pm 0.031$ \\ $\Delta M$ & --- & --- & --- & $0.194$ & $0.193$ & $0.196$ \\
$H_0$ [km s$^{-1}$ Mpc$^{-1}$] & $73.08 \pm 1.00$ & $73.16 \pm 1.03$ & $73.17 \pm 1.03$ & $74.49 \pm 1.10$ & $74.38 \pm 1.13$ & $74.36 \pm 1.10$ \\
$q_0$ & $-0.385 \pm 0.15$ & $-0.393 \pm 0.16$ & $-0.391 \pm 0.15$ & $-0.382 \pm 0.15$ & $-0.392 \pm 0.16$ & $-0.400 \pm 0.16$ \\
$d_{\rm crit}$ [Mpc] & --- & --- & --- & $19.95 \pm 0.1$ & $19.81 \pm 1.04$ & $19.70 \pm 0.96$ \\
\hline
$\chi^2_{\min}$ & 748.46 & --- & --- & 728.26 & --- & --- \\
$\chi^2_{\rm red}$ & 0.91 & --- & --- & 0.89 & --- & --- \\
$\Delta\chi^2$ (vs no-trans) & --- & --- & --- & $-20.20$ & --- & --- \\
$\Delta$AIC & --- & --- & --- & $-16.21$ & --- & --- \\
$\Delta$BIC & --- & --- & --- & $-6.77$ & --- & --- \\
\hline
\textbf{log$\boldsymbol{\mathcal{Z}}$ (Nested)} & --- & --- & $-380.91 \pm 0.20$ & --- & --- & $-377.49 \pm 0.25$ \\
\hline
\multicolumn{7}{c}{$\Delta\log\mathcal{Z}$ (transition $-$ no–transition): \ \ +3.42 $\pm$ 0.32} \\
\hline
\multicolumn{7}{c}{\textit{Bayesian Interpretation:} Moderate evidence for transition model} \\
\hline
\multicolumn{7}{c}{\textit{AIC Interpretation:} Strong preference for transition model \;|\;
\textit{BIC Interpretation:} Moderate preference for transition model} \\
\hline\hline
\end{tabular}
\end{table*}

\begin{table*}[t]
\centering
\caption{\textbf{Flat $w$CDM — Frequentist best-fit $\pm1\sigma$ and Bayesian posterior means $\pm$ 68\% CIs} for no–transition and transition. For Bayesian evidence, $\Delta\log\mathcal{Z}$ is transition $-$ no–transition.}
\label{tab:wcdm_all}
\begin{tabular}{l|ccc|ccc}
\hline\hline
 & \multicolumn{3}{c|}{\textbf{No transition}} & \multicolumn{3}{c}{\textbf{Transition}} \\
\cline{2-7}
\textbf{Parameter} & \textbf{Frequentist} & \textbf{Bayes (MCMC)} & \textbf{Bayes (Nested)} & \textbf{Frequentist} & \textbf{Bayes (MCMC)} & \textbf{Bayes (Nested)} \\
\hline
$M$ & $-19.25 \pm 0.03$ & $-19.247 \pm 0.029$ & $-19.247 \pm 0.029$ & --- & --- & --- \\
$M_<$ & --- & --- & --- & $-19.404 \pm 0.05$ & $-19.400 \pm 0.056$ & $-19.400 \pm 0.054$ \\
$M_>$ & --- & --- & --- & $-19.213 \pm 0.03$ & $-19.211 \pm 0.031$ & $-19.211 \pm 0.032$ \\
$\Delta M$ & --- & --- & --- & $0.191$ & $0.189$ & $0.189$ \\
$H_0$ [km s$^{-1}$ Mpc$^{-1}$] & $73.31 \pm 1.02$ & $73.33 \pm 1.01$ & $73.36 \pm 1.02$ & $74.44 \pm 1.10$ & $74.57 \pm 1.09$ & $74.57 \pm 1.12$ \\
$\Omega_m$ & $0.288 \pm 0.062$ & $0.283 \pm 0.067$ & $0.292 \pm 0.066$ & $0.291 \pm 0.065$ & $0.291 \pm 0.068$ & $0.287 \pm 0.067$ \\
$w$ & $-0.891 \pm 0.15$ & $-0.898 \pm 0.14$ & $-0.919 \pm 0.14$ & $-0.898 \pm 0.13$ & $-0.920 \pm 0.15$ & $-0.912 \pm 0.14$ \\
$d_{\rm crit}$ [Mpc] & --- & --- & --- & $19.90 \pm 0.5$ & $19.74 \pm 1.02$ & $19.64 \pm 0.99$ \\
\hline
$\chi^2_{\min}$ & 1522.48 & --- & --- & 1507.30 & --- & --- \\
$\chi^2_{\rm red}$ & 0.90 & --- & --- & 0.89 & --- & --- \\
$\Delta\chi^2$ (vs no-trans) & --- & --- & --- & $-15.18$ & --- & --- \\
$\Delta$AIC & --- & --- & --- & $-11.18$ & --- & --- \\
$\Delta$BIC & --- & --- & --- & $-0.30$ & --- & --- \\
\hline
\textbf{log$\boldsymbol{\mathcal{Z}}$ (Nested)} & --- & --- & $-772.36 \pm 0.25$ & --- & --- & $-768.69 \pm 0.31$ \\
\hline
\multicolumn{7}{c}{$\Delta\log\mathcal{Z}$ (transition $-$ no–transition): \ \ +3.67 $\pm$ 0.40} \\
\hline
\multicolumn{7}{c}{\textit{Bayesian Interpretation:} Moderate evidence for transition model} \\
\hline
\multicolumn{7}{c}{\textit{AIC Interpretation:} Strong preference for transition model \;|\;
\textit{BIC Interpretation:} Models are statistically indistinguishable} \\
\hline\hline
\end{tabular}
\end{table*}
\end{widetext}
\FloatBarrier

\onecolumngrid
\begin{table*}[t!]
\centering
\caption{\textbf{Flat CPL ($w_0w_a$CDM) — Bayesian posterior means $\pm$ 68\% CIs} for no–transition and transition (Bayesian analysis only). For Bayesian evidence, $\Delta\log\mathcal{Z}$ is transition $-$ no–transition.}
\label{tab:cpl_all}
\begin{tabular}{l|cc|cc}
\hline\hline
 & \multicolumn{2}{c|}{\textbf{No transition}} & \multicolumn{2}{c}{\textbf{Transition}} \\
\cline{2-5}
\textbf{Parameter} & \textbf{Bayes (MCMC)} & \textbf{Bayes (Nested)} & \textbf{Bayes (MCMC)} & \textbf{Bayes (Nested)} \\
\hline
$M$ & $-19.248 \pm 0.029$ & $-19.249 \pm 0.030$ & --- & --- \\
$M_<$ & --- & --- & $-19.402 \pm 0.056$ & $-19.404 \pm 0.057$ \\
$M_>$ & --- & --- & $-19.211 \pm 0.031$ & $-19.211 \pm 0.031$ \\
$\Delta M$ & --- & --- & $0.191$ & $0.193$ \\
$H_0$ [km s$^{-1}$ Mpc$^{-1}$] & $73.31 \pm 1.04$ & $73.28 \pm 1.04$ & $74.57 \pm 1.13$ & $74.58 \pm 1.11$ \\
$\Omega_m$ & $0.306 \pm 0.085$ & $0.298 \pm 0.089$ & $0.314 \pm 0.084$ & $0.295 \pm 0.091$ \\
$w_0$ & $-0.926 \pm 0.14$ & $-0.918 \pm 0.14$ & $-0.936 \pm 0.14$ & $-0.922 \pm 0.14$ \\
$w_a$ & $-0.318 \pm 0.86$ & $-0.256 \pm 0.84$ & $-0.369 \pm 0.86$ & $-0.198 \pm 0.83$ \\
$d_{\rm crit}$ [Mpc] & --- & --- & $19.66 \pm 0.99$ & $19.65 \pm 0.98$ \\
\hline
\textbf{log$\boldsymbol{\mathcal{Z}}$ (Nested)} & --- & $-773.35 \pm 0.29$ & --- & $-769.66 \pm 0.32$ \\
\hline
\multicolumn{5}{c}{$\Delta\log\mathcal{Z}$ (transition $-$ no–transition): \ \ $+3.69 \pm 0.43$} \\
\hline
\multicolumn{5}{c}{\textit{Interpretation:} Moderate evidence for the transition model} \\
\hline\hline
\end{tabular}
\vspace{-5em}
\end{table*}

\begin{figure}[h!]
\centering
\includegraphics[width=1.05\textwidth]{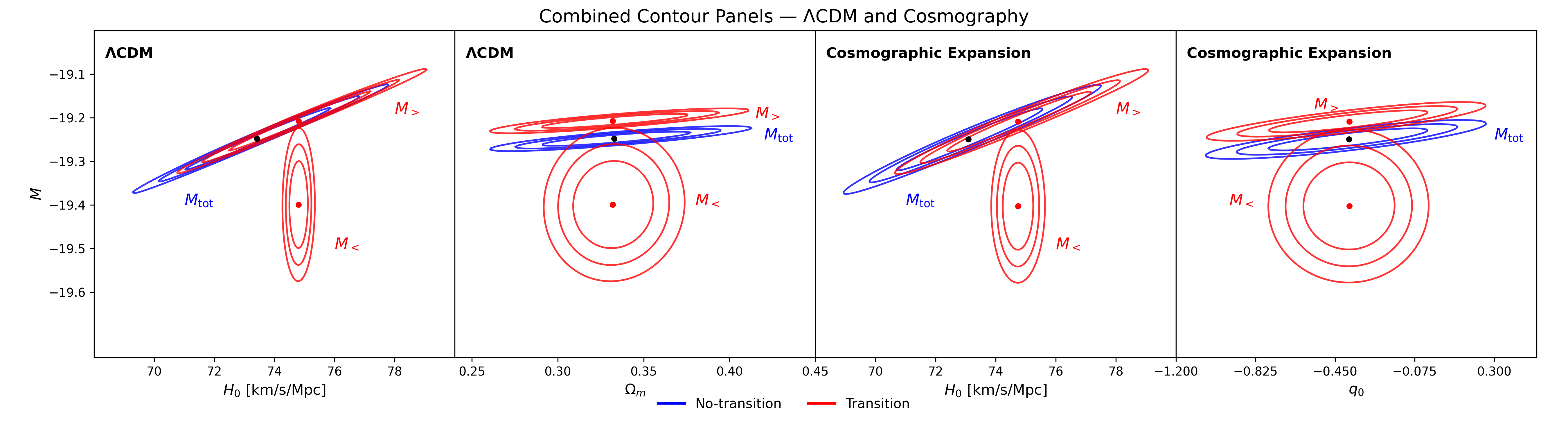}
\caption{\textbf{Frequentist confidence contours for the flat $\Lambda$CDM and cosmographic expansion models.}
Shown are the $1\sigma$, $2\sigma$, and $3\sigma$ confidence regions between $(M, M_{<}, M_{>})$ and $(\Omega_m, H_0, q_0)$.}
\label{fig:lcdm_cosmo_contours}
\end{figure}

\noindent
\begin{minipage}[t]{0.48\textwidth}
\raggedright
Because $d_{\rm crit}$ enters as a threshold parameter, $\chi^2(d_{\rm crit})$ is only piecewise smooth, making its frequentist uncertainty resolution-limited, while the Bayesian posterior provides a more conservative estimate. Reduced $\chi^2$ values slightly below unity are expected due to the conservative Pantheon+ covariance, which includes both statistical and systematic uncertainties without error rescaling. The fact that a consistent transition scale is recovered across all cosmological frameworks considered indicates that this feature is not driven by the specific background parametrization, but instead reflects an empirical preference of the Pantheon+ likelihood itself.
\end{minipage}\hfill
\begin{minipage}[t]{0.48\textwidth}
\raggedright
Across flat $\Lambda$CDM and the cosmographic expansion, the frequentist confidence contours exhibit a consistent pattern: introducing the luminosity transition leads to an increase in $H_0$ of $\sim2\%$, while $\Omega_m$ and $q_0$ remain essentially unchanged. This behavior indicates that the effect is confined to the local distance calibration rather than reflecting a modification of the background expansion. In Fig.~\ref{fig:lcdm_cosmo_contours}, this is directly visible, as $\Omega_m$ and $q_0$ remain stable between the two cases, whereas $H_0$ shifts to higher values when the transition is included.
\end{minipage}

\noindent
\begin{minipage}[t]{0.48\textwidth}
\vspace{2em}
\raggedright
A qualitatively identical behavior is observed in the flat $w$CDM case, as shown in Figs.~\ref{fig:wcdm_overlay} and \ref{fig:wcdm_contours}. The inclusion of the luminosity transition again produces a clear upward shift in the inferred value of $H_0$, while the background cosmological parameters $\Omega_m$ and $w$ remain statistically consistent with their no-transition values. This reinforces the interpretation that the transition acts predominantly as a low-redshift calibration effect, without introducing significant changes to the underlying expansion dynamics.
\end{minipage}\hfill
\begin{minipage}[t]{0.48\textwidth}
\centering
\refstepcounter{figure}
\includegraphics[width=\textwidth]{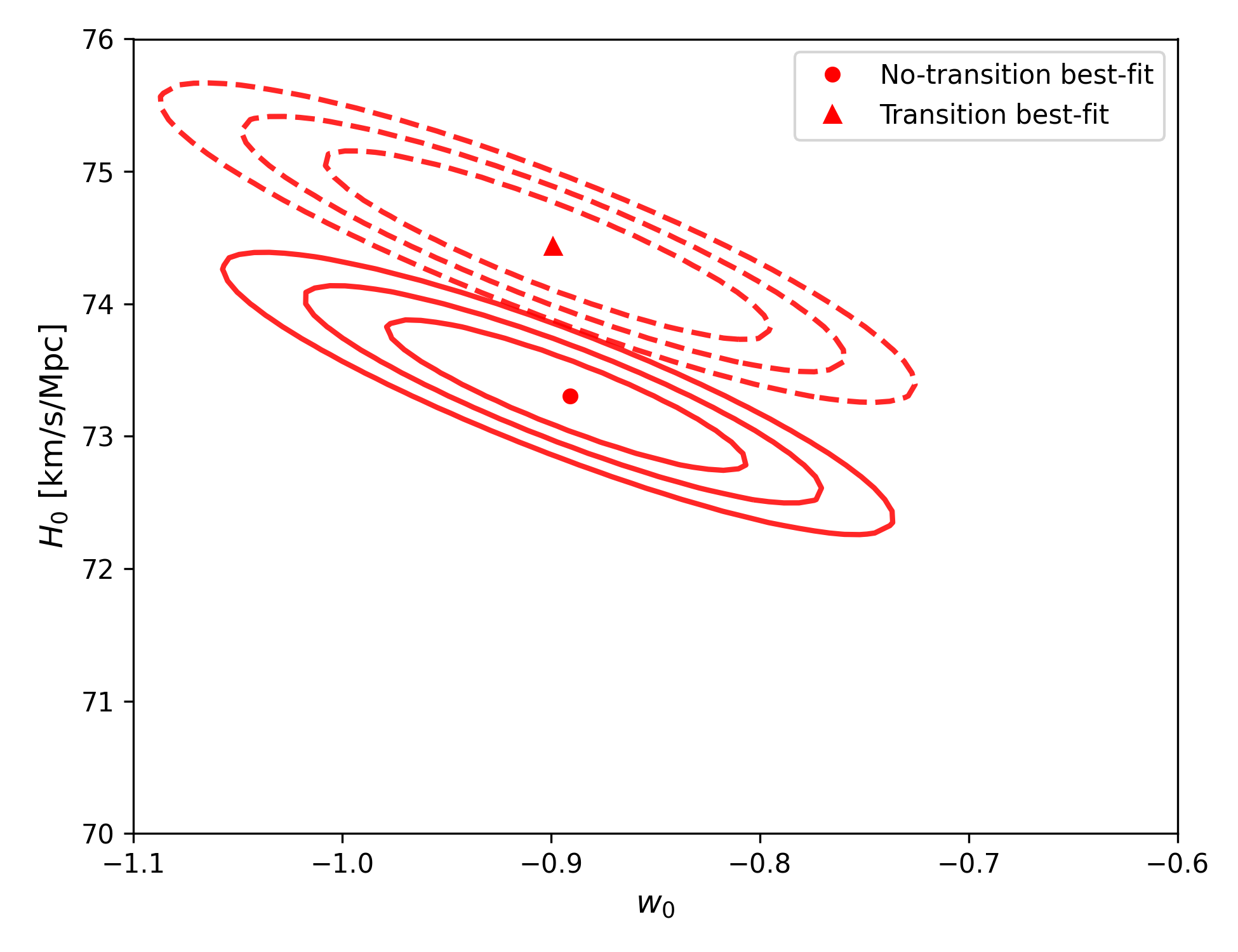}

\vspace{0.4em}
{\small FIG. \thefigure. 1--3$\sigma$ frequentist confidence contours in the $(H_0, w)$ plane for the flat $w$CDM model.}
\label{fig:wcdm_overlay}
\end{minipage}
\vspace{2em}
\begin{figure}[h!]
\centering
\includegraphics[width=0.9\textwidth]{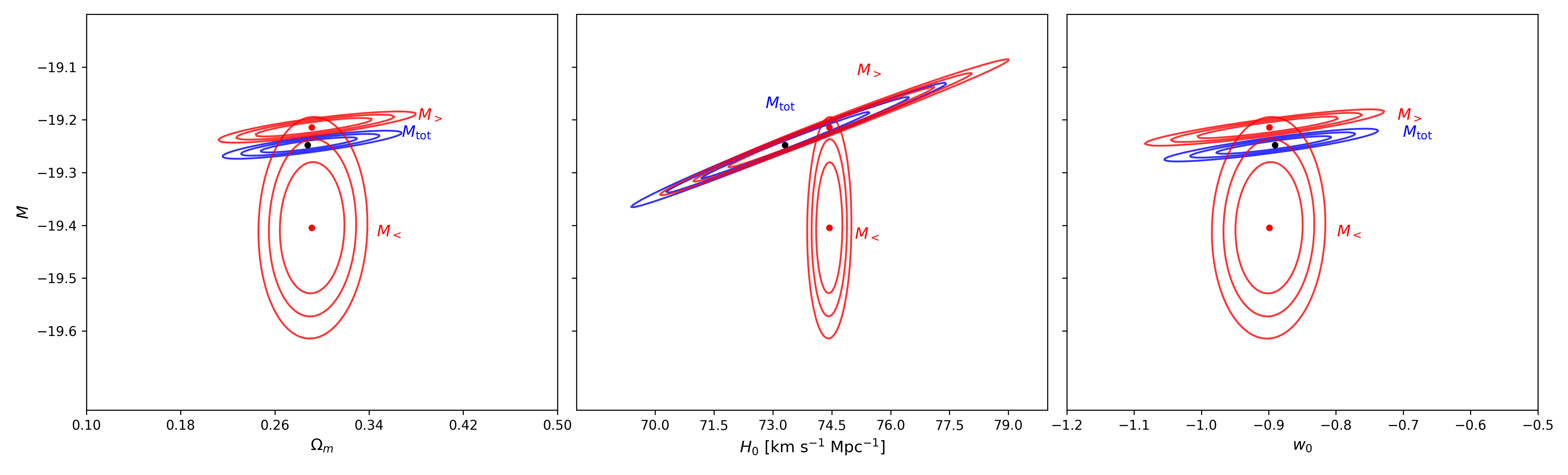}
\caption{\textbf{Frequentist $1\sigma$, $2\sigma$, and $3\sigma$ confidence contours for the flat $w$CDM model.}
The panels show the joint constraints between $(M, M_{<}, M_{>})$ and $(\Omega_m, w, H_0)$, illustrating the effect of the luminosity transition.}
\label{fig:wcdm_contours}
\end{figure}

\begin{figure*}[t!]
\centering
\includegraphics[width=0.90\textwidth]{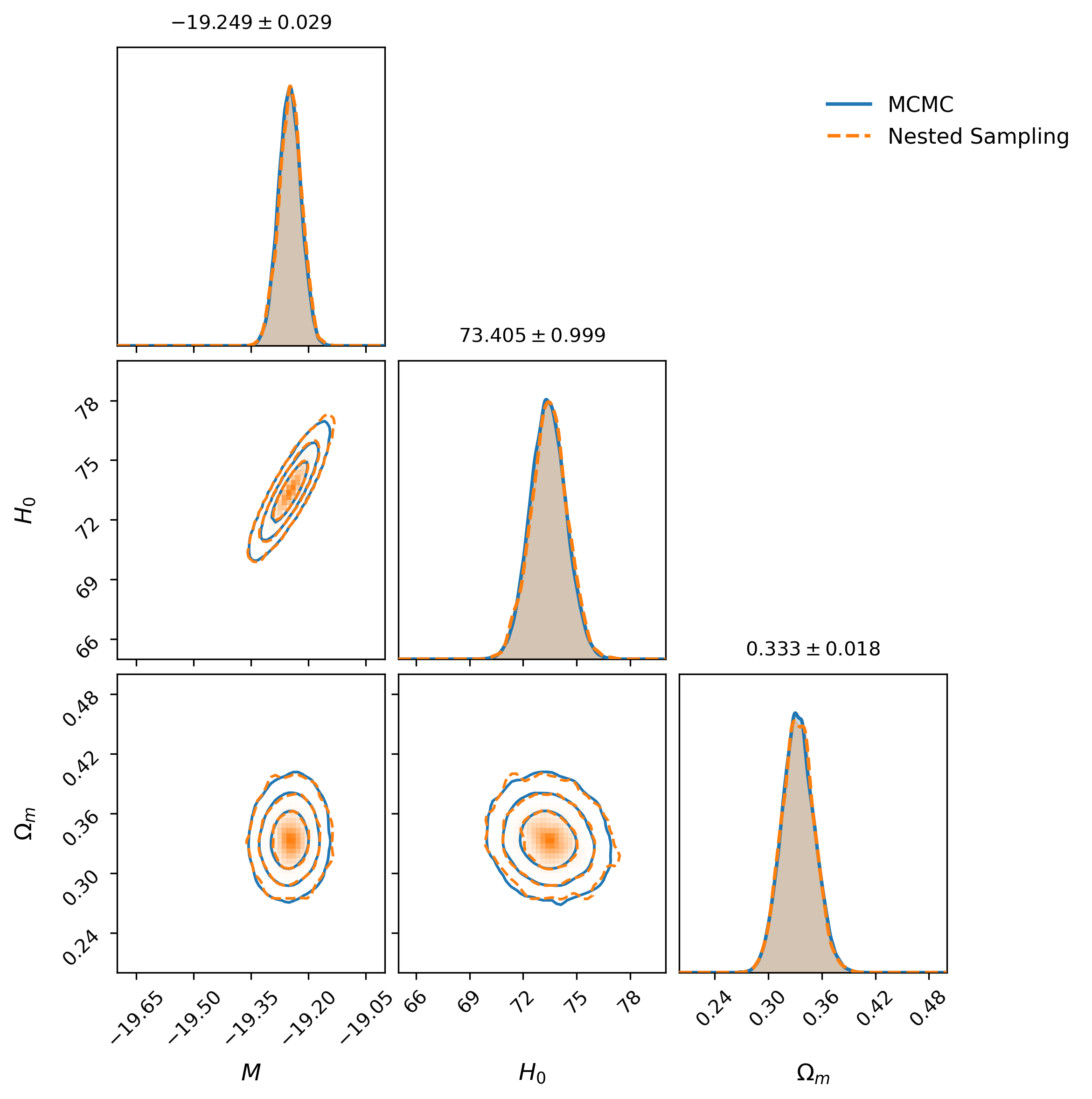}
\caption{\textbf{Flat $\Lambda$CDM — no-transition.}
Overlay of MCMC (blue) and Nested Sampling (orange) posteriors ($1\sigma$ - $3\sigma$).}
\label{fig:corner_lcdm_noT}
\end{figure*}

\begin{figure*}[t!]
\centering
\includegraphics[width=0.96\textwidth]{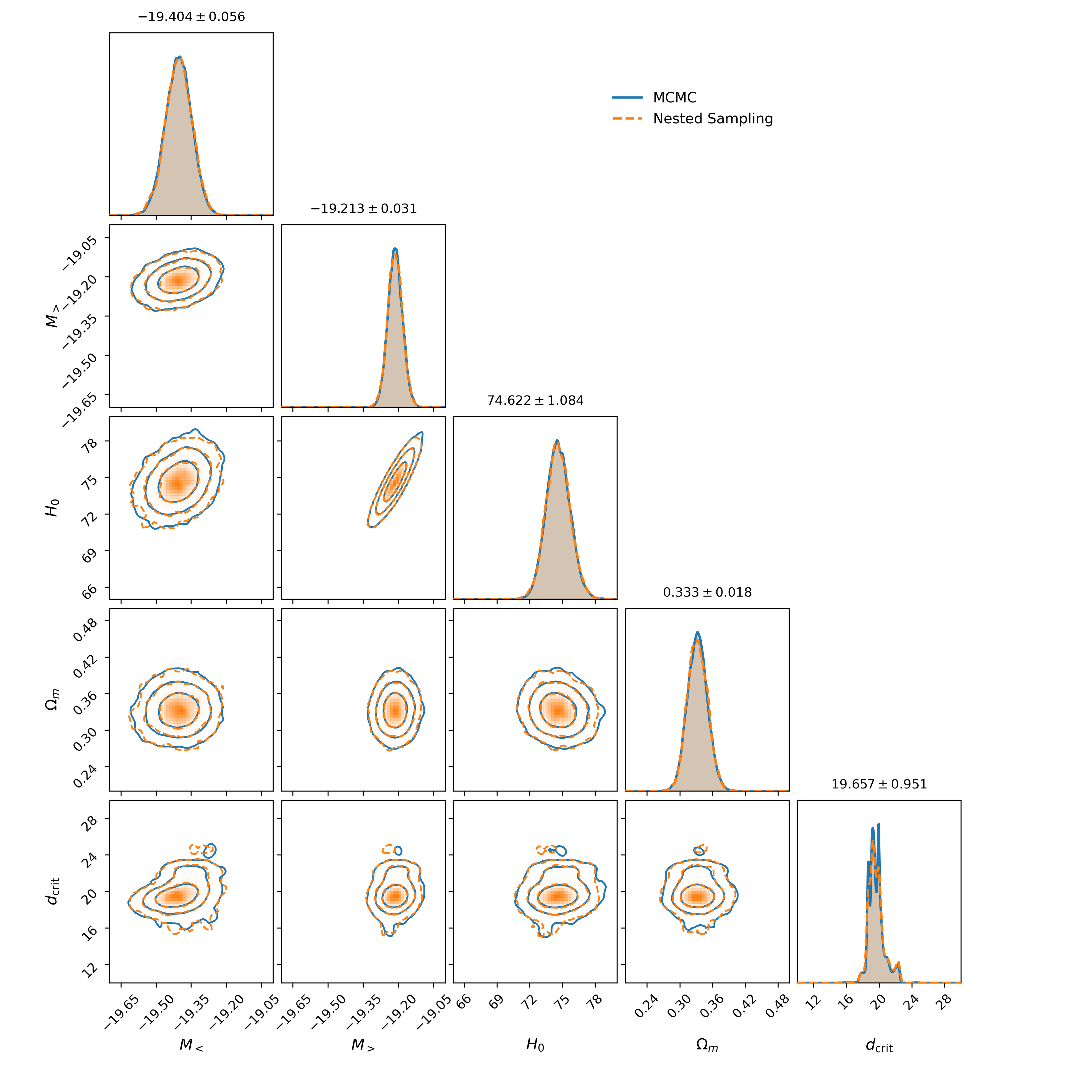}
\caption{\textbf{Flat $\Lambda$CDM — transition.}
Overlay of MCMC (blue) and Nested Sampling (orange) posteriors ($1\sigma$ - $3\sigma$).}
\label{fig:corner_lcdm_T}
\end{figure*}

\vspace{2em}
\noindent
\begin{minipage}[t]{0.48\textwidth}
\raggedright
The Bayesian posterior distributions for the flat $\Lambda$CDM model, for both the no-transition and transition cases, are shown in Figs.~\ref{fig:corner_lcdm_noT} and \ref{fig:corner_lcdm_T}. In each case, we overlay the results obtained from MCMC and Nested Sampling, enabling a direct comparison between the two inference approaches at the level of both the multidimensional parameter contours and the one-dimensional marginalized distributions. The agreement between the two methods confirms the robustness of the inferred constraints, despite their different sampling strategies and computational characteristics.
\end{minipage}\hfill
\begin{minipage}[t]{0.48\textwidth}
\raggedright
The mean parameter values derived from the MCMC chains are indicated on the corresponding one-dimensional posterior distributions. The inclusion of the luminosity transition produces a clear shift in $H_0$, while leaving the remaining parameters largely unaffected, consistent with the frequentist analysis. Since this qualitative behavior is preserved across all cosmological models considered, we present only the $\Lambda$CDM corner plots here for clarity. The corresponding results for the cosmographic, $w$CDM, and CPL models are provided in the Appendix \ref{app:suppfigs} (see, e.g., Fig.~\ref{fig:corner_cosmo_noT}-\ref{fig:corner_cpl_T}).
\end{minipage}

\twocolumngrid
\FloatBarrier

\section{Conclusion–Discussion}\label{Sec.IV}

\textbf{Robustness and Consistency of the Results}. Motivated by the work of Perivolaropoulos and Skara \cite{Perivolaropoulos_2023}, who tested the homogeneity of the Pantheon+ sample within a flat $\Lambda$CDM framework by minimizing a $\chi^2$ likelihood, we extended their analysis to include additional cosmological backgrounds---namely a second--order cosmographic expansion, a flat $w$CDM, and a flat CPL model. Each case besides the CPL model was analyzed using both a frequentist $\chi^2$ minimization and a robust Bayesian inference implemented through Markov Chain Monte Carlo (MCMC) and Nested Sampling. The transition CPL model included 7 parameters to be constrained and a $\chi^2$ minimization could not provide reliable constraints, forcing us to implement only a Bayesian inference for this model via MCMC and Nested Sampling.
Across all tested models, the transition framework is consistently preferred over the corresponding no--transition case, as confirmed by the information criteria (AIC, BIC) and the Bayesian evidence (log~$Z$). The MCMC sampling performed with the \texttt{emcee} ensemble sampler proved computationally lighter and produced stable, symmetric posterior distributions and reliable mean values, however, it cannot by construction calculate the evidence which is needed to perform model comparison. We therefore resorted to Nested Sampling, implemented via \texttt{dynesty}, which enabled accurate estimation of the Bayesian evidence while naturally yielding posterior samples. The slightly skewed contours obtained from Nested Sampling relative to MCMC (see figures \ref{fig:corner_lcdm_noT}-\ref{fig:corner_cpl_T}) are attributed to the smaller number of live points adopted to maintain computational efficiency but do not affect the overall consistency of the inferred constraints.

A potential systematic that could affect low--redshift SNe~Ia analyses is the \textit{volumetric redshift--scatter bias}, which arises from peculiar--velocity and redshift uncertainties combined with the flux--limited nature of nearby surveys. This bias can artificially enhance the brightness of very local supernovae and, if unaccounted for, may mimic a small--scale deviation from homogeneity. As shown by \cite{Perivolaropoulos_2023}, removing non--Cepheid--hosted SNe with $z<0.01$ mitigates this effect, reducing the statistical significance of the transition from $\sim3\sigma$ to $\sim2\sigma$ while preserving its best--fit location near $20~\mathrm{Mpc}$.

\textbf{Interpretation and Physical Implications.} The inferred magnitude offset of $\Delta M = M_{>} - M_{<} \simeq 0.19~\mathrm{mag}$ observed across all models tested corresponds to a systematic increase of the Hubble constant by roughly $2\%$, while the other parameters remain largely unaffected. The consistency of this shift across $\Lambda$CDM, $w$CDM, CPL, and cosmographic frameworks indicates that the phenomenon primarily reflects a calibration--level deviation rather than a change in the underlying cosmological dynamics. The persistence of a critical distance near $d_{\mathrm{crit}} \simeq 20~\mathrm{Mpc}$ suggests a possible link to local--universe structure or subtle astrophysical systematics affecting the nearby calibrator population.

The detection of a transition in the standardized absolute magnitude of Type~Ia supernovae around $d_{\mathrm{crit}} \simeq 20~\mathrm{Mpc}$ has direct implications for the inferred value of the Hubble constant and for the interpretation of the Hubble tension. The best-fit results consistently indicate that nearby supernovae are brighter, with $M_{<} \simeq -19.39$ compared to $M_{>} \simeq -19.20$, corresponding to a luminosity offset of $\Delta M \simeq 0.19~\mathrm{mag}$. If this luminosity difference reflects a real physical transition rather than a statistical fluctuation or local bias, then analyses that assume a single, globally homogeneous $M$ can be biased because the very local calibrator population (Cepheid-hosted SNe) would not be representative of the Hubble-flow population.

In a distance-ladder calibration perspective, an unmodeled offset between the local calibrators and the Hubble-flow sample would propagate into the inferred absolute calibration and hence into the inferred value of $H_0$. In our analysis, where Cepheid-hosted supernovae are incorporated directly into the joint likelihood together with the full Pantheon+ sample and covariance matrix, allowing for two absolute magnitudes $(M_{<},M_{>})$ leads to a modest ($\sim2\%$) increase in the best-fit $H_0$ across all tested models. This behavior indicates that, within the Pantheon+ likelihood framework adopted here, the preferred transition chiefly alters the effective calibration while leaving the background expansion parameters largely unchanged. Overall, if the transition is genuine, it constitutes a localized deviation from homogeneity in the standardized SNe~Ia luminosity that can affect late-time $H_0$ inference and therefore merits further scrutiny as a contributor to the Hubble-tension phenomenology.

Several astrophysical or observational mechanisms could in principle account for this apparent luminosity transition. Differences in progenitor environments between nearby and more distant supernovae may lead to subtle variations in explosion properties even after standardization. Metallicity gradients in the local Universe can affect the nucleosynthetic yields of $^{56}$Ni, thus altering the peak luminosity of SNe~Ia, while small differences in dust properties or reddening laws between Cepheid-hosted and Hubble-flow galaxies could introduce residual color-dependent biases that mimic a magnitude offset. Additional possibilities include mild cross-calibration inconsistencies between low-$z$ photometric systems, imperfect Cepheid distance calibration, or local environmental effects such as bulk flows and peculiar velocities that generate the volumetric redshift scatter bias. Each of these effects could yield a systematic over-luminosity of nearby SNe~Ia without requiring new physics.

However, if such astrophysical or calibration effects cannot fully explain the observed step of $\Delta M \simeq 0.18$~mag, the transition may point to a more fundamental origin. Modified gravity theories that predict a slowly varying effective gravitational constant $G_{\mathrm{eff}}(z)$ can naturally produce a similar luminosity shift \cite{Wright_2018,Marra:2021fvf,Nojiri_2017,astashenok2022chandrasekharmasslimitwhite,Das_2015,Carvalho_2017,Gaztanaga:2001fh,Mould_2014,Ballardini:2020iws,Ballardini_2022,Zhao:2018gwk,Desmond:2019ygn,Jain_2013,Brax:2021wcv,Das_2015,Jain_2016,Alestas:2020zol,Di_Valentino_2024,Perivolaropoulos:2003we,Perivolaropoulos:2005yv,Perivolaropoulos:2019vkb}, since the Chandrasekhar mass and therefore the intrinsic luminosity of SNe~Ia scale as $M_{\mathrm{Ch}} \propto G_{\mathrm{eff}}^{-3/2}$. A small late-time increase in $G_{\mathrm{eff}}$ would enhance the intrinsic brightness of nearby SNe~Ia, generating a local magnitude offset and a corresponding rise in the inferred $H_{0}$. This mechanism, explored in the context of scalar--tensor and screened modified gravity models provides a theoretically motivated framework that can reproduce the observed phenomenology and simultaneously test the constancy of Newton’s constant assumed by $\Lambda$CDM. If confirmed, such a low-redshift deviation in $G_{\mathrm{eff}}$ would place the standard cosmological model under direct observational scrutiny, offering a potential signature of new gravitational physics operating in the local Universe.

In conclusion, our results demonstrate a consistent, cross-model preference for a luminosity transition near $20~\mathrm{Mpc}$ that modestly increases $H_0$ while leaving other cosmological parameters largely unaffected. This behavior points to a localized deviation from complete homogeneity in the SN~Ia calibration and motivates further investigation with next--generation datasets and more comprehensive statistical frameworks.

\newpage

\section*{Data and Code Availability}\label{SecGit}
The numerical files and scripts used for the reproduction of all figures and tables presented in this paper are publicly available in this \href{https://github.com/ChrisStamou/Late-Time-SnIa-Luminosity-Transition-Thesis-2025}{GitHub repository under the MIT license}.

\section*{Acknowledgments}
This research was
supported by COST Action CA21136 - Addressing observational tensions in cosmology with systematics and
fundamental physics (CosmoVerse), supported by COST
(European Cooperation in Science and Technology).
\begin{widetext}
\section*{Appendix}

This Appendix provides supplementary information supporting the statistical interpretation of model comparison tests and the methodological choices used in this analysis.

\subsection*{A. Information Criteria: AIC and BIC}

Model selection within the frequentist framework was guided by the Akaike Information Criterion (AIC) and the Bayesian Information Criterion (BIC), defined respectively as:
\begin{align}
\mathrm{AIC} &= \chi^2_{\min} + 2k , \\
\mathrm{BIC} &= \chi^2_{\min} + k \ln N ,
\end{align}
where $k$ is the number of free parameters and $N$ is the number of data points.  
Lower AIC or BIC values indicate a better model when penalizing excessive parameterization.  
The table below summarizes the interpretation used in this work for the difference $\Delta \mathrm{AIC/BIC} = \mathrm{AIC/BIC}_{\text{model 2}} - \mathrm{AIC/BIC}_{\text{model 1}}$, where negative values favor the transition model.

\begin{table*}[h!]
\centering
\begin{tabular}{|c|l|c|l|}
\hline
$\Delta \mathrm{AIC}$ & Interpretation & $\Delta \mathrm{BIC}$ & Interpretation \\
\hline
$< -10$ & Strong preference for transition model & $< -10$ & Strong preference for transition model \\
\hline
$[-10, -5)$ & Moderate preference for transition model & $[-10, -6)$ & Moderate preference for transition model \\
\hline
$[-5, -2)$ & Weak preference for transition model & $[-6, -2)$ & Weak preference for transition model \\
\hline
$[-2, 2]$ & Models statistically indistinguishable & $[-2, 2]$ & Models statistically indistinguishable \\
\hline
$(2, 5]$ & Weak preference for no-transition model & $(2, 6]$ & Weak preference for no-transition model \\
\hline
$(5, 10]$ & Moderate preference for no-transition model & $(6, 10]$ & Moderate preference for no-transition model \\
\hline
$> 10$ & Strong preference for no-transition model & $> 10$ & Strong preference for no-transition model \\
\hline
\end{tabular}
\caption{Interpretation guide for AIC and BIC differences between the transition and no-transition models.}
\label{tab:aicbic}
\end{table*}

\subsection*{B. Bayesian Evidence and the Jeffreys Scale}

In the Bayesian framework, model comparison is based on the \textit{Bayes factor},
\begin{equation*}
B_{12} = \frac{\mathcal{Z}_1}{\mathcal{Z}_2}, \qquad
\Delta \log \mathcal{Z} = \log \mathcal{Z}_1 - \log \mathcal{Z}_2 ,
\end{equation*}
where $\mathcal{Z}$ denotes the Bayesian evidence for a given model.  
Positive $\Delta \log \mathcal{Z}$ values indicate preference for Model~1 over Model~2.  
Interpretation follows the conventional Jeffreys scale summarized below.

\begin{table}[h!]
\centering
\begin{tabular}{|c|l|}
\hline
$\Delta \log \mathcal{Z}$ & Interpretation \\
\hline
$< 1$ & Inconclusive evidence \\
\hline
$1$ -- $2.5$ & Weak evidence for Model 1 \\
\hline
$2.5$ -- $5$ & Moderate evidence for Model 1 \\
\hline
$> 5$ & Strong evidence for Model 1 \\
\hline
\end{tabular}
\caption{Interpretation of Bayes factors according to the Jeffreys scale.}
\label{tab:jeffreys}
\end{table}


\subsection*{C. Prior Ranges Used in Bayesian Analyses}

All Bayesian runs adopt flat (uniform) priors over the ranges listed in Table~\ref{tab:priors}. 
The adopted intervals were chosen to (i) fully enclose the regions of non-negligible posterior support for each model, and (ii) remain broad enough to avoid artificially truncating parameter degeneracies, particularly in higher-dimensional cases (most notably CPL). 
For this reason, some ranges are model-dependent (e.g.\ wider bounds for $H_0$ and $\Omega_m$ in CPL), reflecting the larger degeneracy volume explored by the sampler rather than any change in the underlying likelihood definition.

Importantly, for all models considered, the resulting marginalized posteriors are well localized within the adopted prior bounds and show no evidence of being prior-edge dominated (i.e.\ the posterior mass does not accumulate near the prior boundaries). 
Thus, the reported parameter constraints are not driven by the imposed limits of the prior ranges.

\begin{table*}[h!]
\centering
\begin{tabular}{l|c}
\hline\hline
Parameter & Flat prior range \\
\hline
$M$ (no-transition) & $[-20,\,-18]$ \\
$M_{<},\,M_{>}$ (transition) & $[-20,\,-18]$ \\
$d_{\rm crit}\,[{\rm Mpc}]$ & $[15,\,25]$ \\
\hline
$H_0\,[{\rm km\,s^{-1}\,Mpc^{-1}}]$ (baseline) & $[60,\,80]$ \\
$H_0$ (cosmography/$w$CDM) & $[65,\,78]$ \\
$H_0$ (CPL) & $[60,\,85]$ \\
\hline
$\Omega_m$ (baseline/CPL) & $[0.1,\,0.5]$ \\
$\Omega_m$ ($w$CDM) & $[0.1,\,0.8]$ \\
\hline
$q_0$ (cosmography) & $[-1,\,0]$ \\
$w$ (constant-$w$) & $[-1.3,\,0]$ \\
$w_0$ (CPL) & $[-2,\,0]$ \\
$w_a$ (CPL) & $[-2,\,2]$ \\
\hline\hline
\end{tabular}
\caption{Prior ranges adopted in the Bayesian analyses. All priors are uniform within the stated intervals.}
\label{tab:priors}
\end{table*}

\subsection*{D. Redshift Cut-off in Cosmographic Expansion}

To ensure the validity of the second-order cosmographic expansion,
\[
d_L(z) = \frac{c}{H_0}\left[z + \frac{1}{2}(1 - q_0)z^2\right],
\]
a maximum redshift $z_{\max}$ must be chosen such that higher-order ($\mathcal{O}(z^3)$) terms remain negligible while still retaining a statistically meaningful number of SNe~Ia.
We therefore tested several redshift cut-offs, $z_{\max} \in \{0.05,\,0.10,\,0.15,\,0.18\}$, for both the no-transition and transition cosmographic models.
The corresponding best-fit parameters are shown below.

\begin{table}[h!]
\centering
\begin{tabular}{c|ccc|c}
\hline\hline
$z_{\max}$ & $M$ & $H_0$ [km s$^{-1}$ Mpc$^{-1}$] & $q_0$ & $\chi^2$ \\
\hline
0.05 & $-19.25075 \pm 0.02941$ & $72.75 \pm 1.23$ & $-0.07720 \pm 0.69218$ & 600.29 \\
0.10 & $-19.25008 \pm 0.02941$ & $73.10 \pm 1.10$ & $-0.40530 \pm 0.34609$ & 691.24 \\
0.15 & $-19.24963 \pm 0.02940$ & $73.09 \pm 1.04$ & $-0.38445 \pm 0.16103$ & 748.46 \\
0.18 & $-19.24865 \pm 0.02940$ & $72.84 \pm 1.03$ & $-0.16950 \pm 0.11733$ & 823.50 \\
\hline\hline
\end{tabular}
\caption{Best-fit parameters for increasing redshift cut-offs using the $q_0$ parametrization (no-transition model).}
\label{tab:zmax_notrans}
\end{table}

\begin{table}[h!]
\centering
\begin{tabular}{c|ccccc|c}
\hline\hline
$z_{\max}$ & $M_{<}$ & $M_{>}$ & $H_0$ [km s$^{-1}$ Mpc$^{-1}$] & $q_0$ & $d_{\mathrm{crit}}$ [Mpc] & $\chi^2$ \\
\hline
0.05 & $-19.40510 \pm 0.04674$ & $-19.20891 \pm 0.03086$ & $74.18 \pm 1.29$ & $-0.10000 \pm 0.69314$ & $19.95 \pm 0.10$ & 579.62 \\
0.10 & $-19.40280 \pm 0.04668$ & $-19.20863 \pm 0.03086$ & $74.49 \pm 1.16$ & $-0.40165 \pm 0.34593$ & $19.95 \pm 0.10$ & 670.99 \\
0.15 & $-19.40222 \pm 0.04666$ & $-19.20821 \pm 0.03086$ & $74.49 \pm 1.11$ & $-0.38340 \pm 0.16101$ & $19.95 \pm 0.10$ & 728.26 \\
0.18 & $-19.40082 \pm 0.04664$ & $-19.20734 \pm 0.03086$ & $74.22 \pm 1.10$ & $-0.17002 \pm 0.11733$ & $19.95 \pm 0.10$ & 803.41 \\
\hline\hline
\end{tabular}
\caption{Best-fit parameters for the transition model with varying redshift cut-offs.}
\label{tab:zmax_trans}
\end{table}

\vspace{0.4cm}
At very low redshifts ($z_{\max}\!\leq\!0.10$), the uncertainties on the deceleration parameter $q_0$ are extremely large (e.g.\ $\sigma_{q_0}\!\approx\!0.69$ for $z_{\max}\!=\!0.05$), reflecting the limited number of nearby SNe~Ia and the dominance of local peculiar–velocity scatter.  
As $z_{\max}$ increases, the statistical precision improves, with $z_{\max}\!=\!0.15$ yielding well–constrained values of $q_0$ while remaining within the validity range of the second–order expansion.  
For $z_{\max}\!>\!0.15$, higher–order terms become non–negligible, leading to increasing $\chi^2$ and biased estimates of $q_0$.

The transition model yields consistently lower $\chi^2$ values and a stable critical distance of $d_{\mathrm{crit}}\!\approx\!20$~Mpc, independent of the redshift cut-off, indicating that the inferred luminosity–magnitude transition is robust.  
Therefore, $z_{\max}=0.15$ was adopted as the optimal cut-off scale in the cosmographic analysis, balancing statistical precision, model validity, and consistency across the no–transition and transition frameworks.

\subsection*{E. Supplementary Corner Plots}\label{app:suppfigs}

For conciseness, the main text shows the representative $\Lambda$CDM plots together with the frequentist $w$CDM contours, while the remaining cosmographic, $w$CDM corner, and CPL plots are collected here.

\begin{figure*}[h!]
\centering
\begin{minipage}[t]{0.48\textwidth}
\centering
\includegraphics[width=\linewidth]{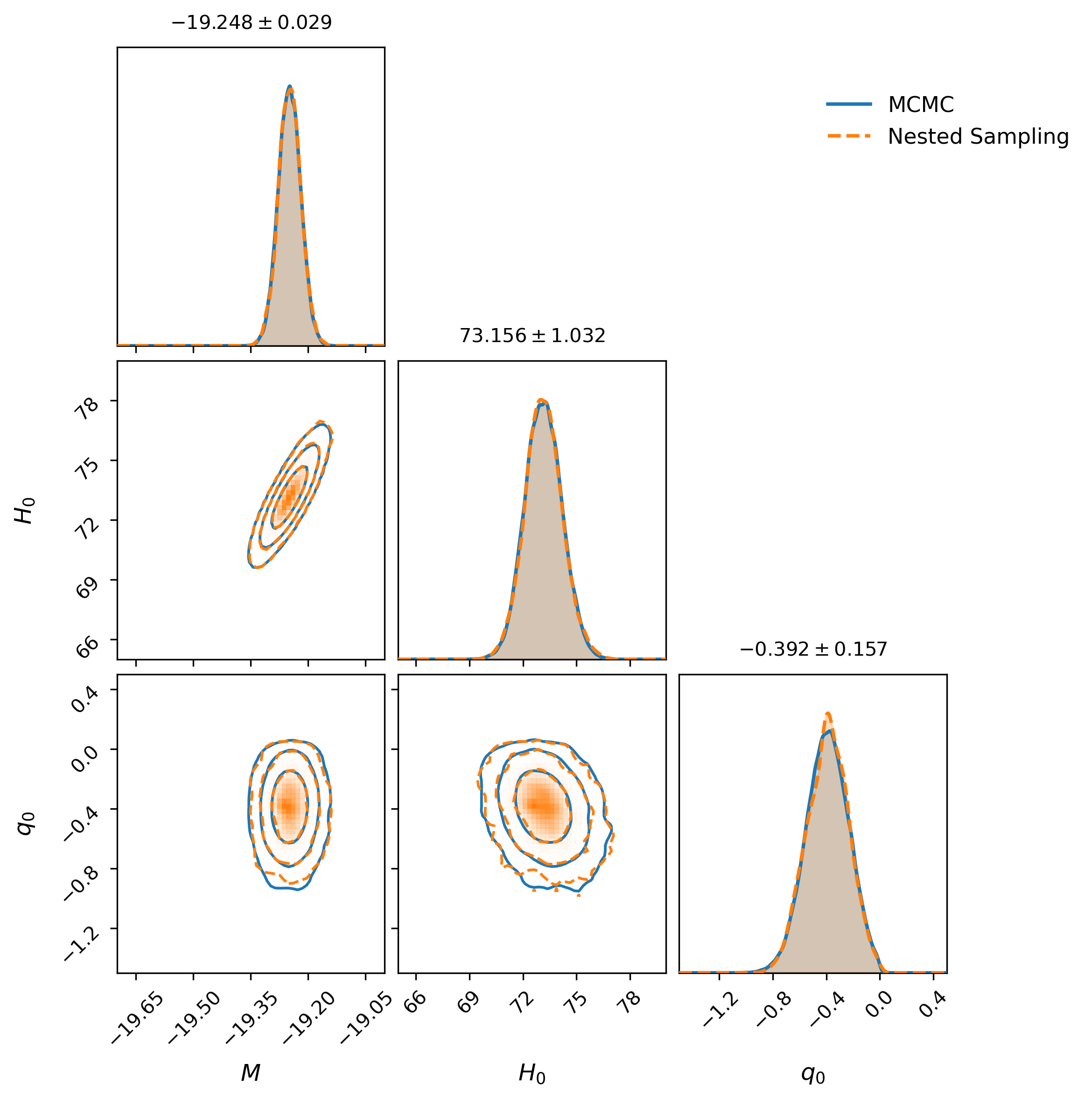}
\caption{\textbf{Cosmographic expansion (2nd order) — no-transition.}
Overlay of MCMC (blue) and Nested Sampling (orange) posteriors ($1\sigma$ - $3\sigma$).}
\label{fig:corner_cosmo_noT}
\end{minipage}\hfill
\begin{minipage}[t]{0.48\textwidth}
\centering
\includegraphics[width=1.16\linewidth,clip,trim=0 0 0 0]{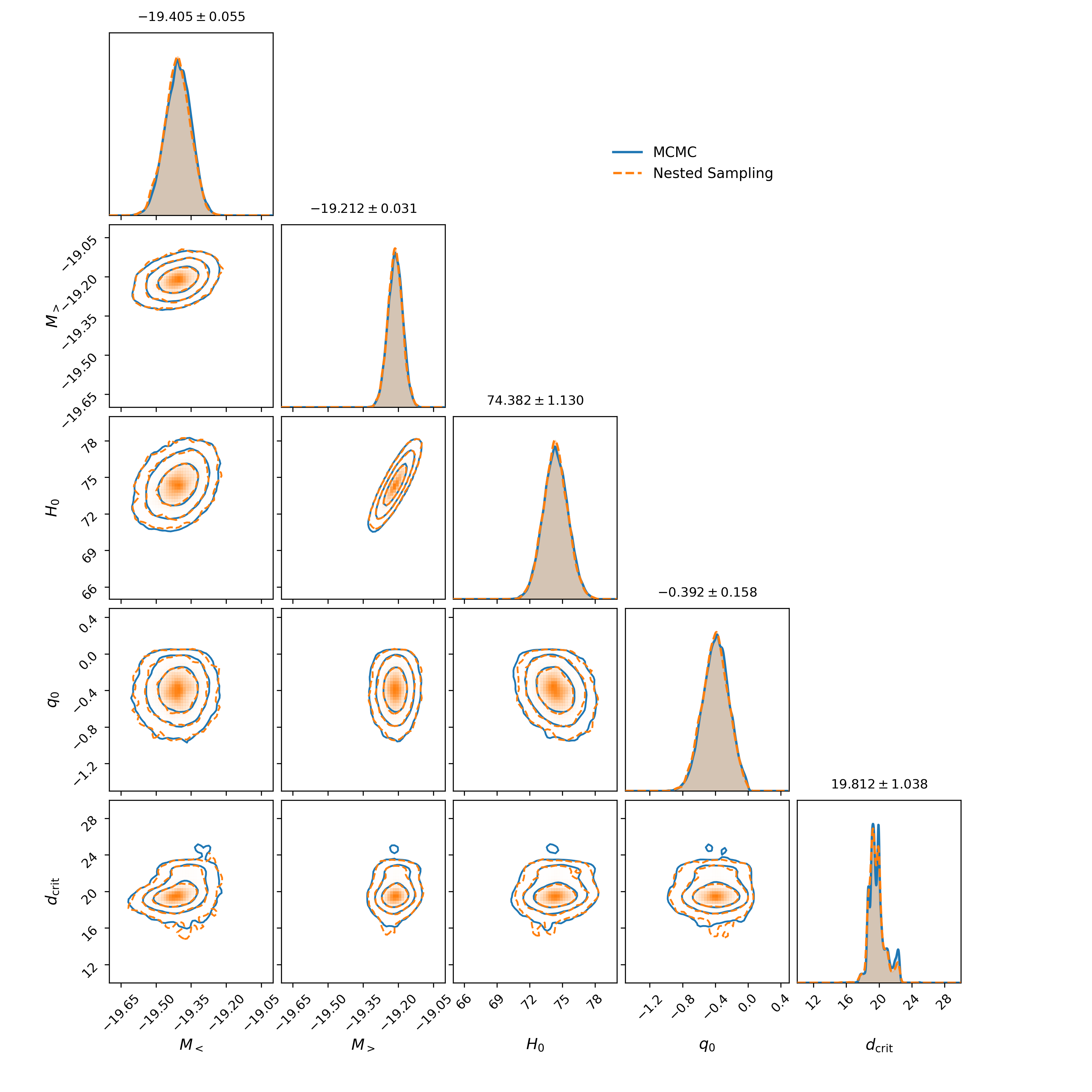}
\caption{\textbf{Cosmographic expansion (2nd order) — transition.}
Overlay of MCMC (blue) and Nested Sampling (orange) posteriors ($1\sigma$ - $3\sigma$).}
\label{fig:corner_cosmo_T}
\end{minipage}
\end{figure*}

\begin{figure*}[h!]
\centering
\begin{minipage}[t]{0.48\textwidth}
\centering
\includegraphics[width=\linewidth]{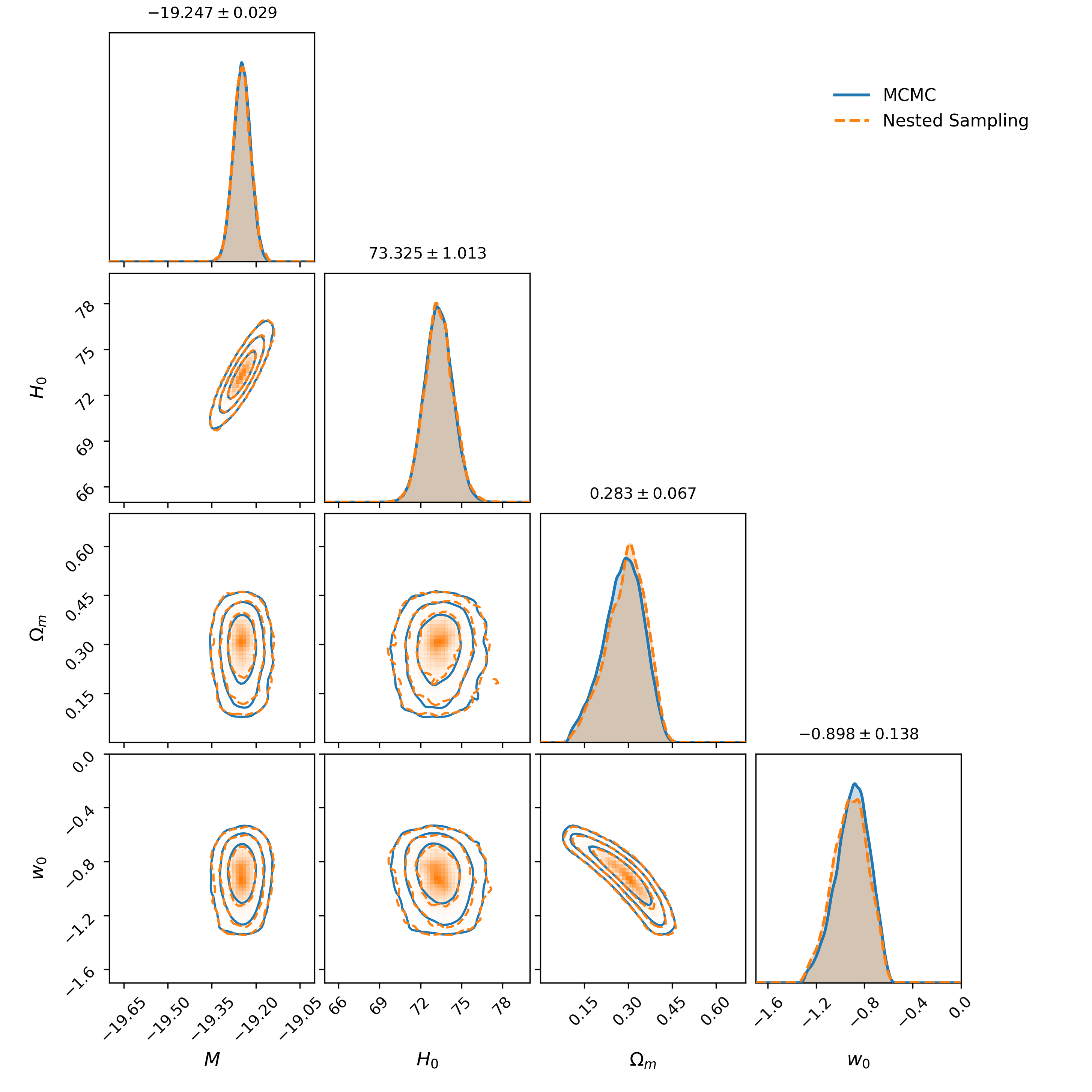}
\caption{\textbf{Flat $w$CDM — no-transition.}
Overlay of MCMC (blue) and Nested Sampling (orange) posteriors ($1\sigma$ - $3\sigma$).}
\label{fig:corner_wcdm_noT}
\end{minipage}\hfill
\begin{minipage}[t]{0.48\textwidth}
\centering
\includegraphics[width=1.20\linewidth,clip,trim=0 0 0 0]{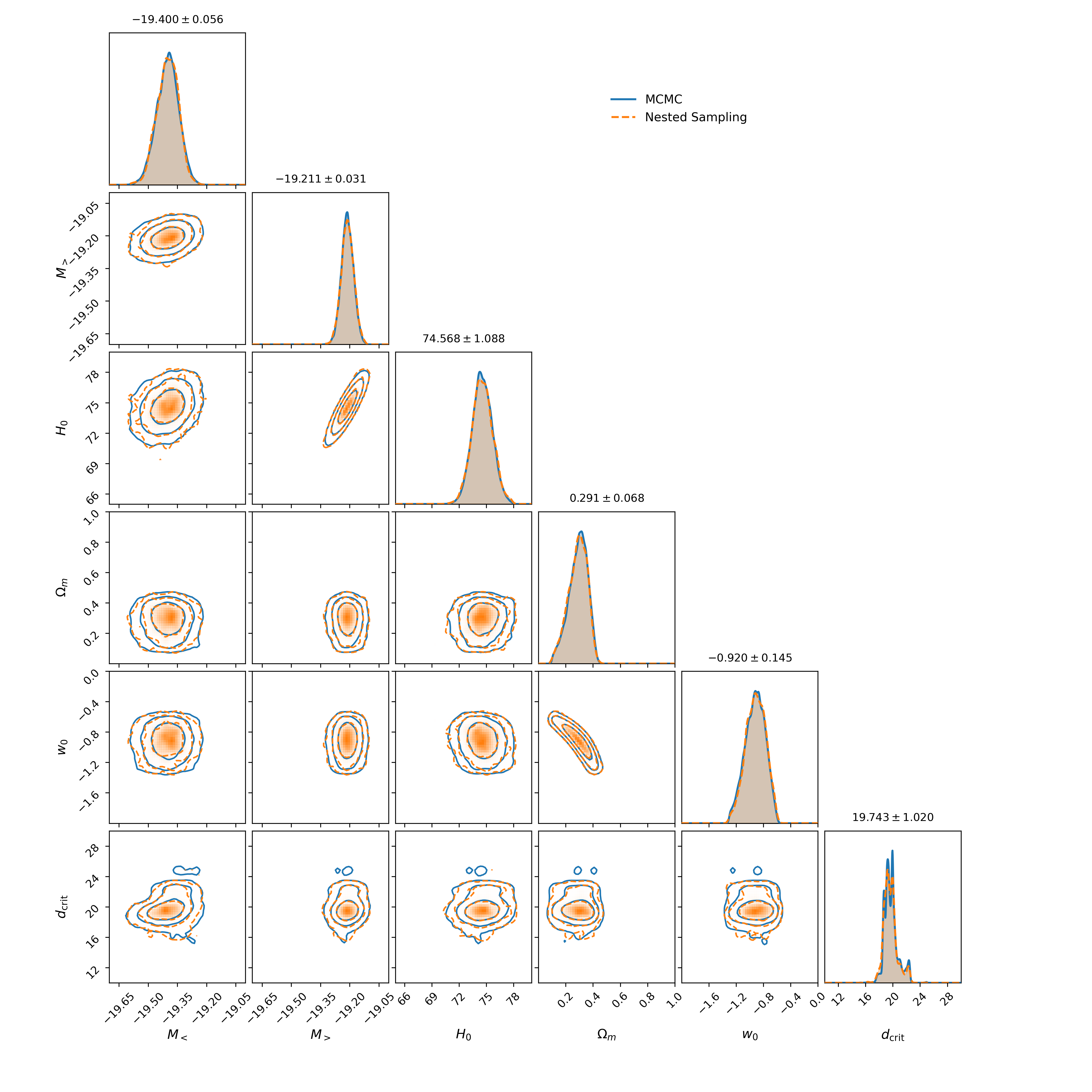}
\caption{\textbf{Flat $w$CDM — transition.}
Overlay of MCMC (blue) and Nested Sampling (orange) posteriors ($1\sigma$ - $3\sigma$).}
\label{fig:corner_wcdm_T}
\end{minipage}
\end{figure*}

\begin{figure*}[h!]
\centering
\includegraphics[width=1.06\textwidth,height=0.47\textheight,keepaspectratio,clip]{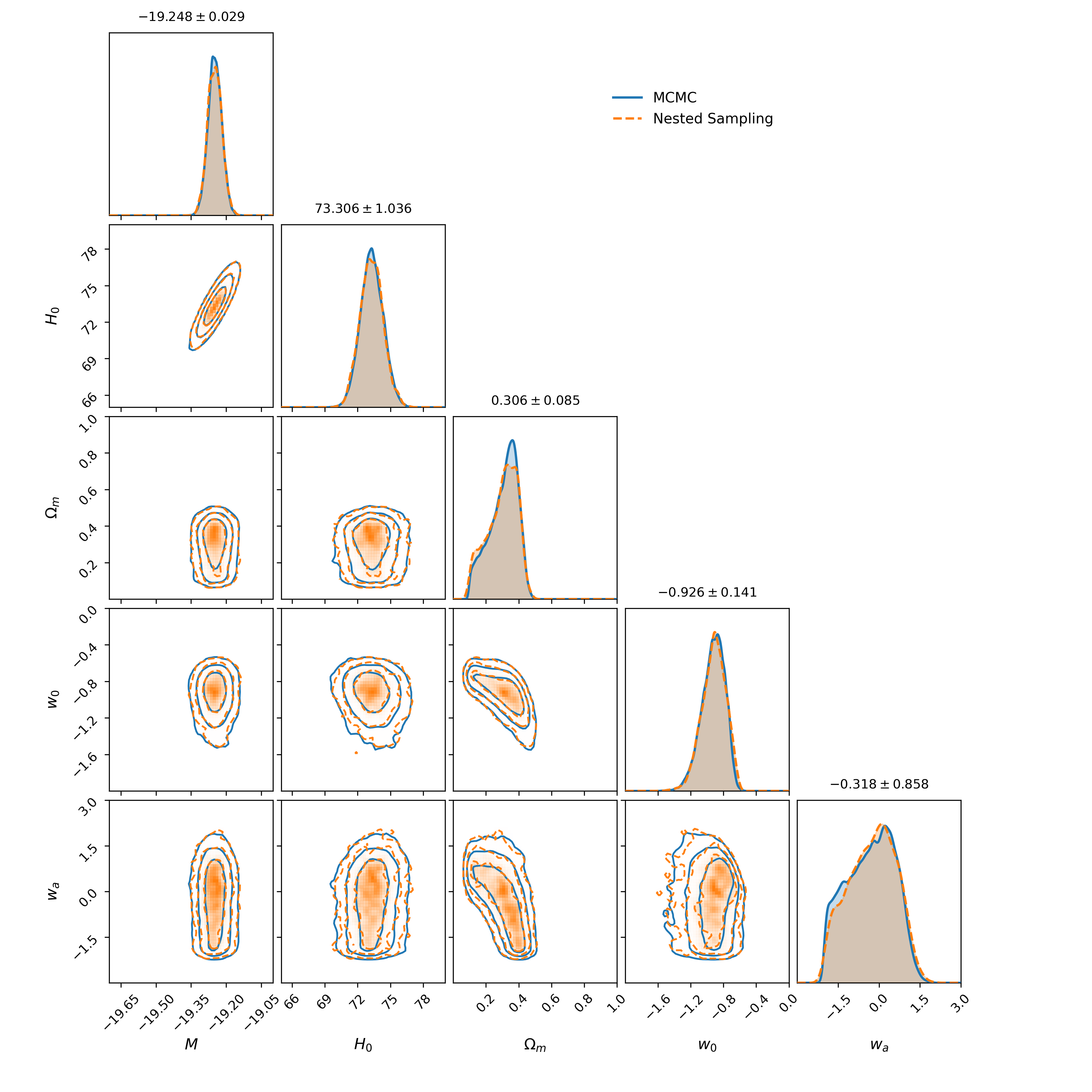}
\caption{\textbf{Flat CPL ($w_0w_a$CDM) — no-transition.}
Overlay of MCMC (blue) and Nested Sampling (orange) posteriors ($1\sigma$ - $3\sigma$).}
\label{fig:corner_cpl_noT}
\end{figure*}

\begin{figure*}[h!]
\centering
\includegraphics[width=1.25\textwidth,height=0.47\textheight,keepaspectratio,clip]{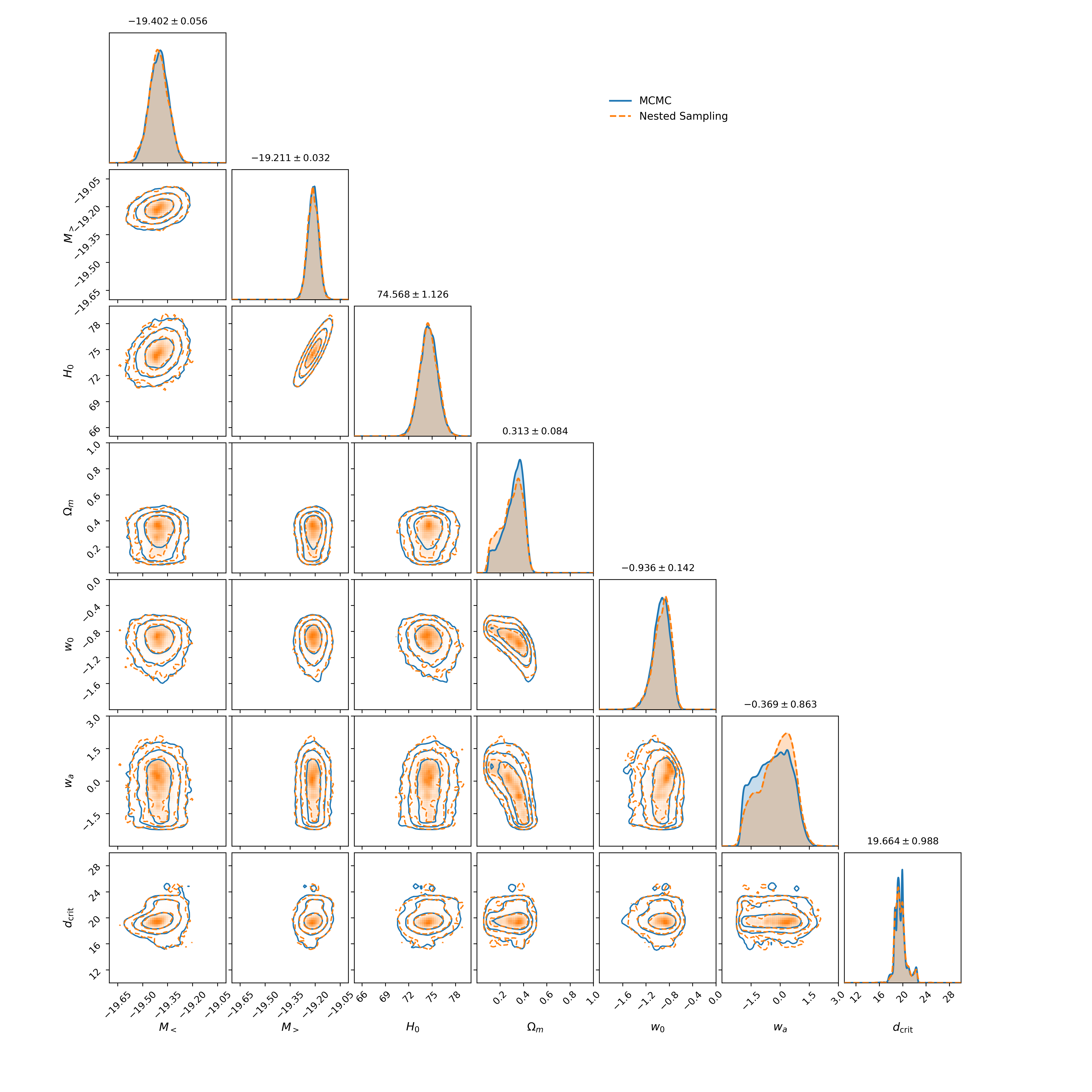}
\caption{\textbf{Flat CPL ($w_0w_a$CDM) — transition.}
Overlay of MCMC (blue) and Nested Sampling (orange) posteriors ($1\sigma$ - $3\sigma$).}
\label{fig:corner_cpl_T}
\end{figure*}

\end{widetext}
\clearpage

\twocolumngrid
\raggedleft
\bibliography{Citations_Biblio}
\end{document}